\documentclass[10pt, conference]{IEEEtran}
\usepackage[titlenumbered, ruled, linesnumbered]{algorithm2e}
\usepackage{microtype, booktabs, amsmath, color, url}
\usepackage{graphicx, tabularx, subfigure, multirow, balance}

\let\oldnl\nl
\newcommand{\nonl}{\renewcommand{\nl}{\let\nl\oldnl}}
\begin{document}

\title{Protecting real-time GPU kernels on integrated CPU-GPU SoC platforms}
\author{Waqar Ali, Heechul Yun\\
University of Kansas, USA.\\
\{wali, heechul.yun\}@ku.edu\\
}

\maketitle

\begin{abstract}
Integrated CPU-GPU architecture provides excellent acceleration capabilities for
data parallel applications on embedded platforms while meeting the size, weight
and power (SWaP) requirements. However, sharing of main memory between CPU
applications and GPU kernels can severely affect the execution of GPU kernels
and diminish the performance gain provided by GPU. For example, in the NVIDIA
Jetson TX2 platform, an integrated CPU-GPU architecture, we observed that, in
the worst case, the GPU kernels can suffer as much as 3X slowdown in the
presence of co-running memory intensive CPU applications. In this paper, we
propose a software mechanism, which we call BWLOCK++, to protect the performance
of GPU kernels from co-scheduled memory intensive CPU applications.
\end{abstract}

\section{Introduction} \label{intro}
Graphic Processing Units (GPUs) are increasingly important computing resources
to accelerate a growing number of data parallel applications. In recent years,
GPUs have become a key requirement for intelligent and timely processing of
large amount of sensor data in many robotics applications, such as UAVs and autonomous
cars. These intelligent robots are, however, resource constrained real-time
embedded systems that not only require high computing performance but also must
satisfy a variety of constraints such as size, weight, power
consumption (SWaP) and cost.
This makes integrated CPU-GPU architecture based computing
platforms, which integrate CPU and GPU in a single chip (e.g., NVIDIA's
Jetson~\cite{jetson} series), an appealing solution for such robotics
applications because of their high performance and
efficiency~\cite{nathan2017a}.
\begin{figure}[t]
  \centering
  \includegraphics[width=0.95\linewidth, height=7cm]{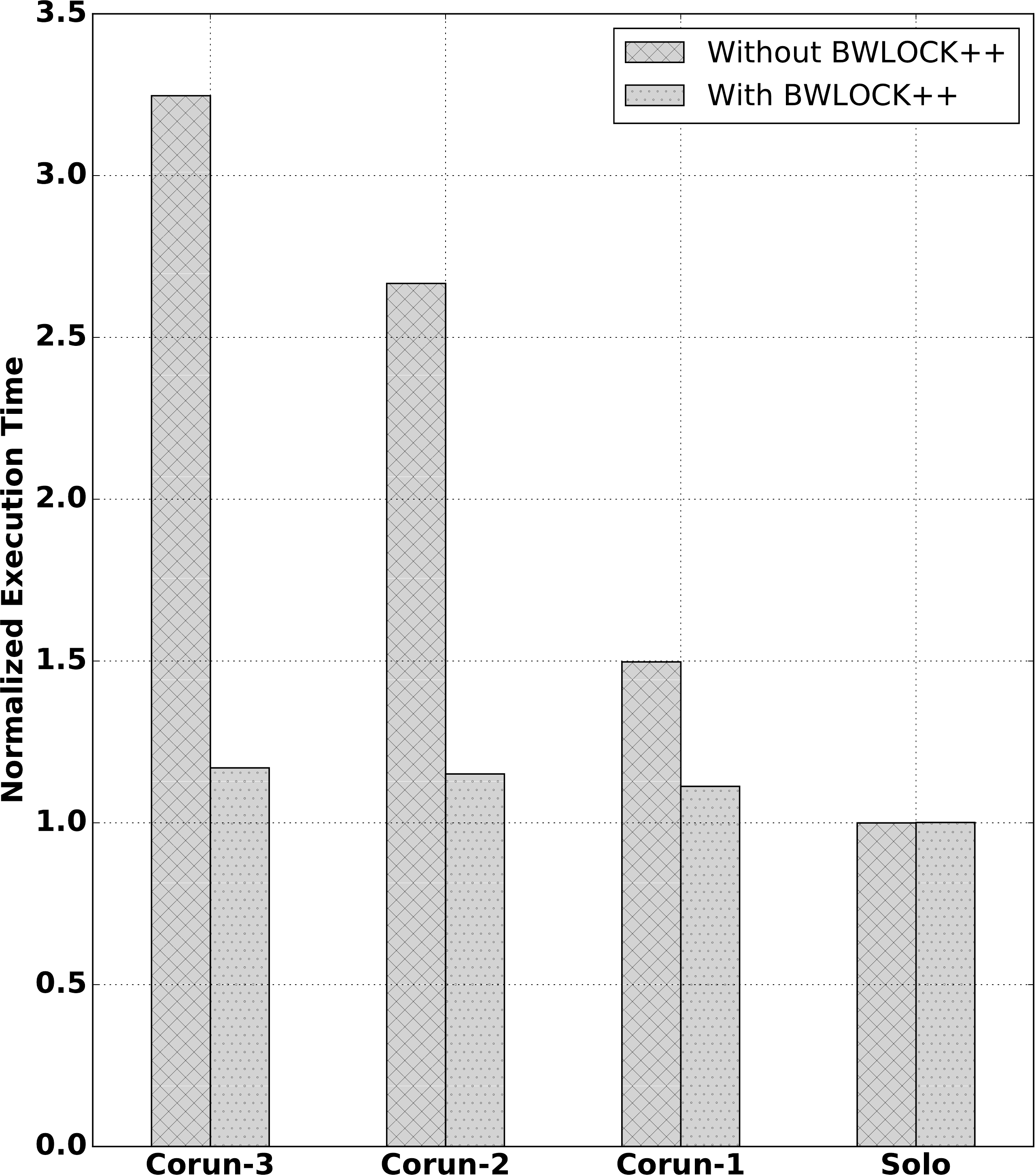}
  \caption{Performance of \emph{histo} benchmark on NVIDIA Jetson TX2 with CPU corunners}
  \label{fig:DNN}
\end{figure}

% specific challenge
Designing critical real-time applications on integrated CPU-GPU architectures
is, however, challenging because contention in the shared hardware
resources (e.g., memory bandwidth) can significantly alter the
applications' timing characteristics. On an integrated CPU-GPU platform, such as
NVIDIA Jetson TX2, the CPU cores and the GPU typically share a single main
memory subsystem. This allows memory intensive batch jobs running on
the CPU cores to significantly interfere with the execution of
critical real-time GPU tasks (e.g., vision based navigation and
obstacle detection) running in parallel due to memory bandwidth
contention.

To illustrate the significance of the problem stated above, we evaluate the
effect of co-scheduling memory bandwidth intensive synthetic CPU benchmarks on the
performance of a GPU benchmark \emph{histo} from the parboil benchmark
suite~\cite{parboil} on a NVIDIA Jetson TX2 platform (See
Table~\ref{tbl:profile} in Section~\ref{sec:eval} for the detailed
time breakdown of \emph{histo}.)
%% The \emph{histo} benchmark's execution time is dominated by the GPU
%% kernel execution time as most of the work is offloaded to the GPU.
%% We measure the performance of the benchmark in terms of its total execution
%% which is comprised of the GPU kernel execution time, CPU compute time and time
%% for copying data between CPU and GPU. 
%% In case of \emph{histo} benchmark, the
%% total execution time is completely dominated by the GPU kernel execution time.
We first run the benchmark alone and record the solo execution
statistics. We then repeat the experiment with an increasing number of
interfering memory intensive benchmarks on the idle CPU cores to observe their
impact on the performance of the \emph{histo} benchmark with and without the
BWLOCK++ framework, which we propose in this paper.
Figure~\ref{fig:DNN} shows
the results of this experiment. As can be seen in \emph{`Without BWLOCK++'},
co-scheduling the memory intensive tasks on the idle CPU cores significantly
increase the execution time of the GPU benchmark---a 3.3X increase---despite the
fact that the benchmark has exclusive access to the GPU. The main cause of the
problem is that, in the Jetson TX2 platform, both CPU and GPU share the main
memory and its limited memory bandwidth becomes a bottleneck. As a result, even
though the platform offers plenty of raw performance, no real-time execution
guarantees can be provided if the system is left unmanaged. In \emph{`With
BWLOCK++'}, on the other hand, performance of the GPU benchmark remains close to
its solo performance measured in isolation.

% our approach
BWLOCK++ is a software framework designed to mitigate the memory bandwidth
contention problem in integrated CPU-GPU architectures. More specifically, we
focus on protecting real-time GPU tasks from the interference of non-critical
but memory intensive CPU tasks. BWLOCK++ dynamically instruments GPU tasks at
run-time and inserts a \emph{memory bandwidth lock} while critical GPU kernels
are being executed on the GPU. When the bandwidth lock is being held by the GPU,
the OS throttles the maximum memory bandwidth usage of the CPU cores to a
certain threshold value to protect the GPU kernels. The threshold value is
determined on a per GPU task basis and may vary depending on the GPU task's
sensitivity to memory bandwidth contention. Throttling CPU cores inevitably
negatively affects the CPU throughput. To minimize the throughput impact, we
propose a throttling-aware CPU scheduling algorithm, which we call Throttle Fair
Scheduler (TFS). TFS favors CPU intensive tasks over memory intensive ones while
the GPU is busy executing critical GPU tasks in order to minimize CPU
throttling. Our evaluation shows that BWLOCK++ can provide good performance
isolation for bandwidth intensive GPU tasks in the presence of memory intensive
CPU tasks. Furthermore, the TFS scheduling algorithm reduces the CPU throughput
loss by up to $75\%$. Finally, we show how BLWOCK++ can be incorporated in
existing CPU focused real-time analysis frameworks to analyze schedulability of
real-time tasksets, utilizing both CPU and GPU.

In this paper, we make the following contributions:
\begin{itemize}
\item We apply memory bandwidth throttling to the problem of
  protecting GPU accelerated real-time tasks from memory intensive CPU
  tasks on integrated  CPU-GPU architecture
\item We identify a negative feedback effect of memory bandwidth
  throttling when used with Linux's CFS~\cite{cfslinux} scheduler. We
  propose a throttling-aware CPU scheduling algorithm, which we call
  Throttle Fair Scheduler (TFS), to mitigate the problem
\item We introduce an automatic GPU kernel instrumentation method that
  eliminates the need of manual programmer intervention to protect GPU
  kernels
\item We implement the proposed framework, which we call BWLOCK++, on
  a real platform, NVIDIA Jetson TX2, and present detailed evaluation
  results showing practical benefits of the framework~\footnote{The
    source code of BWLOCK++ is publicly available at: \url{https://github.com/wali-ku/BWLOCK-GPU}}
\item We show how the proposed framework can be integrated into the
  existing CPU focused real-time schedulability analysis framework
\end{itemize}

The remainder of this paper is organized as follows. We present necessary
background and discuss related work in Section~\ref{sec:background}. In
Section~\ref{sec:model}, we present our system model. Section~\ref{sec:bwlock++}
describes the design of our software framework BWLOCK++ and
Section~\ref{sec:implementation} presents implementation details. In
Section~\ref{sec:eval}, we describe our evaluation 
platform and present evaluation results using a set of GPU benchmarks. In
Section~\ref{sec:analysis}, we present the analysis framework of
BWLOCK++ based real-time systems. We discuss limitations of our approach in
Section~\ref{sec:discussion} and conclude in Section~\ref{sec:conclusion}.

\section{Background and Related Work}\label{sec:background}
In this section, we provide necessary background and discuss related work.

GPU is an accelerator that executes some specific functions requested by a
master CPU program. Requests to the GPU can be made by using GPU programming
frameworks such as CUDA that offer standard APIs. A request to GPU is typically
composed of the following four predictable steps:
\begin{itemize}
\item Copy data from host memory to device (GPU) memory
\item Launch the function---called \emph{kernel}---to be executed on the GPU
\item Wait until the kernel finishes
\item Copy the output from device memory to host memory
\end{itemize}

In the real-time systems community, GPUs have been studied actively in recent
years because of their potential benefits in accelerating demanding
data-parallel real-time applications~\cite{Kato2015}. As observed
in~\cite{agarwal2015}, GPU kernels typically demand high memory bandwidth to
achieve high data parallelism and, if the memory bandwidth required by GPU
kernels is not satisfied, it can result in significant performance reduction.
For discrete GPUs, which have dedicated graphic memories, researchers have
focused on addressing interference among the co-scheduled GPU tasks. Many
real-time GPU resource management frameworks adopt scheduling based approaches,
similar to real-time CPU scheduling, that provide priority or server based
scheduling of GPU tasks~\cite{Kato2011, Kato2012, zhou2015}. Elliot et al.,
formulate the GPU resource management problem as a synchronization problem and
propose the GPUSync framework that uses real-time locking protocols to
deterministically handle GPU access requests~\cite{elliot2013}. Here, at any
given time, one GPU kernel is allowed to utilize the GPU to eliminate the
unpredictability caused by co-scheduled GPU kernels. In
~\cite{kim2017rtcsa-server}, instead of using a real-time locking protocol that
suffers from busy-waiting at the CPU side, the authors propose a GPU server
mechanism which centralizes access to the GPU and allows CPU suspension (thus
eliminating the CPU busy-waiting). All the aforementioned frameworks primarily
work for discrete GPUs, which have dedicated graphic memory, but they do not
guarantee predictable GPU timing on integrated CPU-GPU based platforms because
they do not consider the problem of the shared memory bandwidth contention
between the CPU and the GPU.

Integrated GPU based platforms have recently gained much attention in the
real-time systems community. In ~\cite{nathan2017a, nathan2017b}, the authors
investigate the suitability of NVIDIA's Tegra X1 platform for use in safety
critical real-time systems. With careful reverse engineering, they have
identified undisclosed scheduling policies that determine how concurrent GPU
kernels are scheduled on the platform. In SiGAMMA ~\cite{sigamma2017}, the
authors present a novel mechanism to preempt the GPU kernel using a
high-priority spinning GPU kernel to protect critical real-time CPU
applications. Their work is orthogonal to ours as it solves the problem of
protecting CPU tasks from GPU tasks while our work solves the problem of
protecting GPU tasks from CPU tasks.

More recently, GPUGuard~\cite{forsberg2017} provides a mechanism for
deterministically arbitrating memory access requests between CPU cores and GPU
in heterogeneous platforms containing integrated GPUs. They extend the PREM
execution model~\cite{pellizzoni2011}, in which a (CPU) task is assumed to have
distinct computation and memory phases, to model GPU tasks. GPUGuard provides
deterministic memory access by ensuring that only a single PREM memory phase is
in execution at any given time. Although GPUGuard can provide strong isolation
guarantees, the drawback is that it may require significant restructuring of
application source code to be compatible with the PREM model.

In this paper, we favor a less intrusive approach that requires minimal or no
programmer intervention. Our approach is rooted on a kernel-level memory
bandwidth throttling mechanism called MemGuard~\cite{yun2013rtas}, which
utilizes hardware performance counters of the CPU cores to limit memory
bandwidth consumption of the individual cores for a fixed time interval on
homogeneous multicore architectures. MemGuard enables a system designer---not
individual application programmers---to partition memory bandwidth among the CPU
cores. However, MemGuard suffers from system-level throughput reduction due to
its coarse-grain bandwidth control (per-core-level control). In contrast,
~\cite{yun2017bwlock} is also based on a memory bandwidth throttling mechanism
on homogeneous multicore architectures but it requires a certain degree of
programmer intervention for fine-grain bandwidth control by exposing a simple
lock-like API to applications. The API can enable/disable memory bandwidth
control in a fine-grain manner within the application source code. However, this
means that the application source code must be modified to leverage the feature.

Our work is based on memory bandwidth throttling, but, unlike prior throttling
based approaches, focuses on the problem of protecting GPU accelerated real-time
tasks on integrated CPU-GPU architectures and does not require any programmer
intervention. Furthermore, we identify a previously unknown negative
side-effect of memory bandwidth throttling when used with Linux's CFS scheduler,
which we mitigate in this work. In the following, we start by defining the
system model, followed by detailed design and implementation of the proposed
system.

\section{System Model}\label{sec:model}
We assume an integrated CPU-GPU architecture based platform, which is composed
of multiple CPU cores and a single GPU that share the same main memory
subsystem. We consider independent periodic real-time tasks with implicit
deadlines and best-effort tasks with no real-time constraints.

{\bf Task Model.} Each task is composed of at least one CPU execution segment
and zero or more GPU execution segments. We assume that \emph{GPU execution is
non-preemptible} and we do not allow concurrent execution of multiple GPU
kernels from different tasks at the same time. Simultaneously co-scheduling
multiple kernels is called GPU co-scheduling, which has been avoided in most
prior real-time GPU management
approaches~\cite{Kato2012,elliot2013,kim2017rtcsa-server} as well due to
unpredictable timing. According to ~\cite{nathan2017a}, preventing GPU
co-scheduling does not necessarily hurt---if not improve---performance because
concurrent GPU kernels from different tasks are executed in a time-multiplexed
manner rather than being executed in parallel.~\footnote{Another recent
study~\cite{amert2017} finds that GPU kernels can only be executed in parallel
if they are submitted from a single address space. In this work, we
assume that a task has its own address space, whose GPU kernels are
thus time-multiplexed with other tasks' GPU kernels at the GPU-level.}

Executing GPU kernels typically requires copying considerable amount of data
between the CPU and the GPU. In particular, synchronous copy directly
contributes to the task's execution time, while asynchronous copy can overlap
with GPU kernel execution. Therefore, we model synchronous copy separately.
Lastly, we assume that a task is single-threaded with respect to the CPU. Then,
we can model a real-time task as follows:
$$\tau_i := \left( C_i, G_i^m, G_i^e, P_i \right)$$
\noindent where:
\begin{itemize}
\item $C_i$ is the cumulative WCET of CPU-only execution
\item $G_i^m$ is the cumulative WCET of synchronous memory operations between
CPU and GPU
\item $G_i^e$ is the cumulative WCET of GPU kernels
\item $P_i$ is the period
\end{itemize}

Note that the goal of BWLOCK++ is to reduce $G^m_i$ and $G^e_i$ under the
presence of memory intensive best-effort tasks running in parallel.

{\bf CPU Scheduling.} We assume a fixed-priority preemptive real-time scheduler
is used for scheduling real-time tasks and a virtual run-time based fair sharing
scheduler (e.g., Linux's Completely Fair Scheduler~\cite{cfslinux}) is used for
best-effort tasks. For simplicity, we assume a single dedicated real-time core
schedules all real-time tasks, while any core can schedule best-effort tasks.
Because GPU kernels are executed serially on the GPU, as mentioned above, for
GPU intensive real-time tasks, which we focus on in this work, this assumption
does not significantly under-utilize the system, especially when there are
enough co-scheduled best-effort tasks, while it enables simpler analysis.

\section{BWLOCK++}\label{sec:bwlock++}
In this section, we provide an overview of BWLOCK++ and discuss its design details.
\begin{figure} [t]
\centering
  \centering
  \includegraphics[width=0.95\linewidth, height = 7cm]{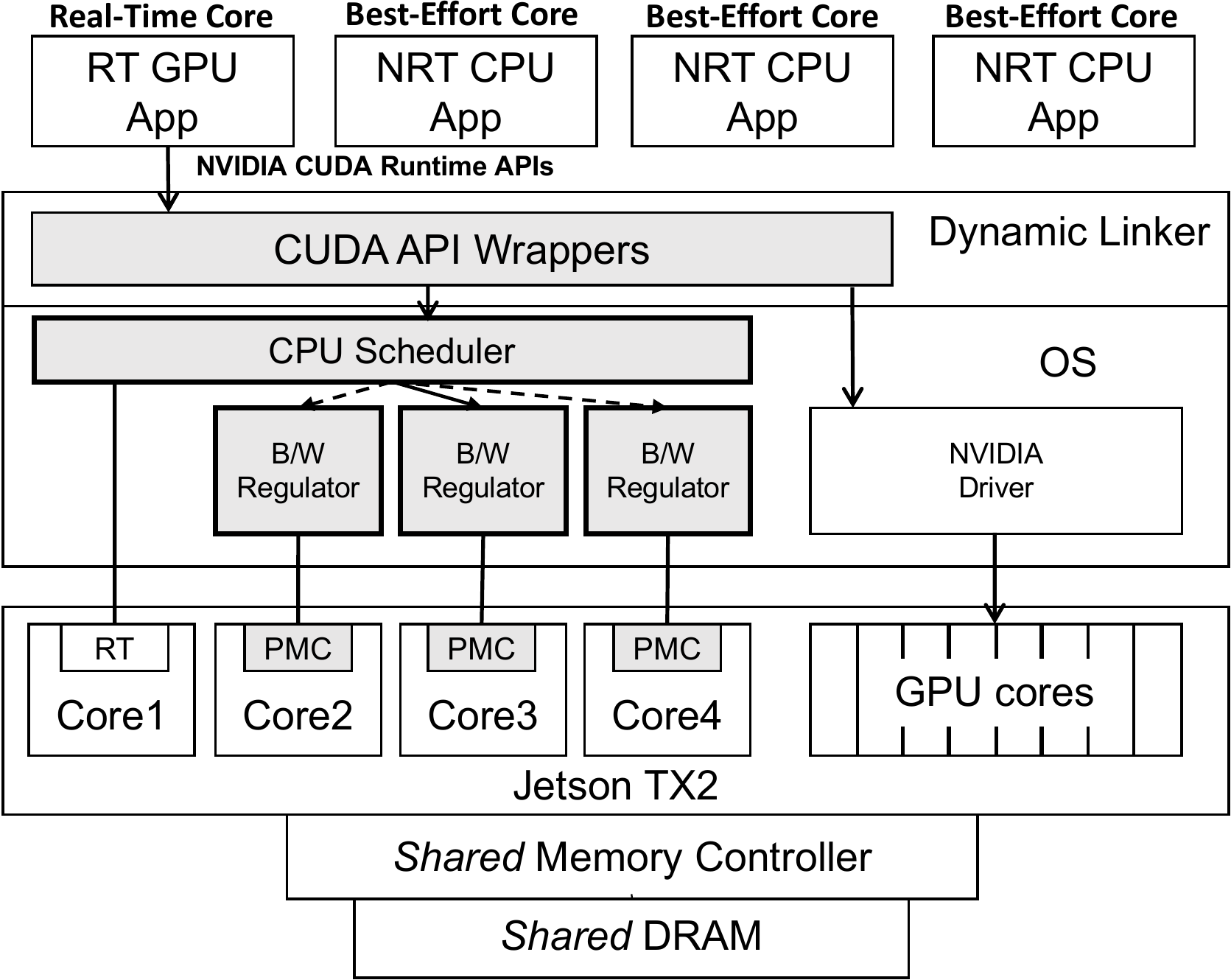}
  \caption{BWLOCK++ System Architecture}
  \label{fig:architecture}
\end{figure}

\subsection{Overview}
BWLOCK++ is a software framework to protect GPU applications on integrated
CPU-GPU architecture based SoC platforms. We focus on the problem of the shared
memory bandwidth contention between GPU kernels and CPU tasks in integrated
CPU-GPU architectures. More specifically, we focus on protecting GPU execution
intervals of real-time GPU tasks from the interference of non-critical but
memory intensive CPU tasks.

In BWLOCK++, we exploit the fact that each GPU kernel is executed via explicit
programming interfaces from a corresponding host CPU program. In other words, we
can precisely determine when the GPU kernel starts and finishes by instrumenting
these functions.

To avoid memory bandwidth contention from the CPU, we notify the OS before a GPU
application launches a GPU kernel and after the kernel completes with the help
of a system call. Apart from acquiring the bandwidth lock on the task's behalf,
this system call also implements the priority ceiling
protocol~\cite{sha1990priority} to prevent preemption of the GPU using task.
While the bandwidth lock is being held by the GPU task, the OS regulates memory
bandwidth consumption of the best-effort CPU cores to minimize bandwidth
contention with the GPU kernel. Concretely, each best-effort core is
periodically given a certain amount of memory bandwidth budget. If the core uses
up its given budget for the specified period, the (non-RT) CPU tasks running on
that core are throttled. In this way, the GPU kernel suffers minimal memory
bandwidth interference from the best-effort CPU cores. However, throttling CPU
cores can significantly lower the overall system throughput. To minimize the
negative throughput impact, we propose a new CPU scheduling algorithm, which we
call the Throttle Fair Scheduler (TFS), to minimize the duration of CPU
throttling without affecting memory bandwidth guarantees for real-time GPU
applications.

Figure~\ref{fig:architecture} shows the overall architecture of the BWLOCK++
framework on an integrated CPU-GPU architecture (NVIDIA Jetson TX2 platform).
BWLOCK++ is comprised of three major components: (1) Dynamic run-time library
for instrumenting GPU applications; (2) the Throttle Fair Scheduler; (3)
Per-core B/W regulator. Working together, they protect real-time GPU kernels and
minimize CPU throughput reduction. We will explain each component in the
following sub-sections.
\begin{figure} [t]
\centering
  \includegraphics[width=0.95\linewidth, height = 6cm]{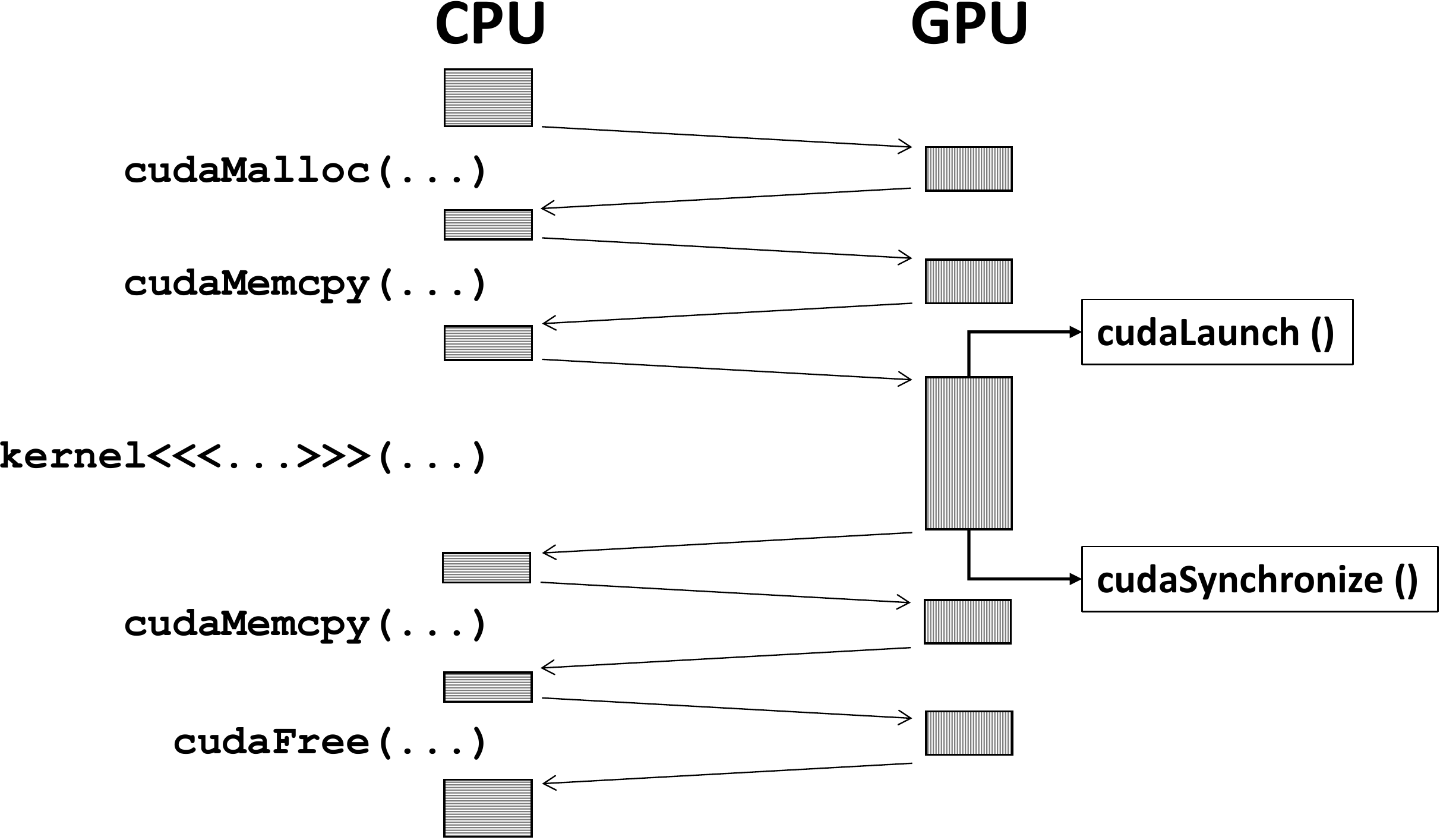}
  \caption{Phases of GPU Application under CUDA Runtime}
  \label{fig:ld_preload}
\end{figure}

\subsection{Automatic Instrumentation of GPU Applications} \label{sec:ld-preload}
To eliminate manual programming efforts, we automatically instrument the program
binary at the dynamic linker level. We exploit the fact that the execution of a
GPU application using a GPU runtime library such as NVIDIA CUDA typically
follows fairly predictable patterns. Figure~\ref{fig:ld_preload} shows the
execution timeline of a typical synchronous GPU application that uses the CUDA
API.
\begin{table*}[t]
	\centering
	\begin{tabularx}{\textwidth}{l| p{7.0cm} |p{7.0cm}}
		\toprule
		API			& Action									& Description \\
		\midrule
		cudaConfigureCall	& Update active streams								& Specify the launch parameters for the CUDA kernel \\
		cudaMemcpy		& Acquire BWLOCK++ (\emph{Before}) Release BWLOCK++ (\emph{After}) 		& Perform synchronous memory copy between CPU and GPU \\
		cudaMemcpyAsync		& Acquire BWLOCK++ and update active streams 					& Perform asynchronous memory copy between CPU and GPU \\
		cudaLaunch		& Acquire BWLOCK++ 								& Launch a GPU kernel \\
		cudaDeviceSynchronize	& Release BWLOCK++ and clear active streams 					& Block the calling CPU thread until all the previously requested tasks in a specific GPU device have completed \\
		cudaThreadSynchronize	& Release BWLOCK++ and clear active streams 					& Deprecated version of cudaDeviceSynchronize \\
		cudaStreamSynchronize   & Update active streams and release BWLOCK++ if there are no active streams 	& Block the calling CPU thread until all the previously requested tasks in a specific CUDA stream have completed \\
		\bottomrule
	\end{tabularx}
	\caption{CUDA APIs instrumented via \texttt{LD\_PRELOAD} for BWLOCK++}
	\label{tbl:cuda-apis}
\end{table*}

In order to protect the runtime performance of a GPU application from co-running
memory intensive CPU applications, we need to ensure that the GPU application
automatically holds the memory bandwidth lock while a GPU kernel is executing on
the GPU or performing a memory copy operation between CPU and GPU. Upon the
completion of the execution of the kernel or memory copy operation, the GPU
application again shall automatically release the bandwidth lock. This is done
by instrumenting a small subset of CUDA API functions that are invoked when
launching or synchronizing with a GPU kernel or while performing a memory copy
operation. These APIs are documented in Table~\ref{tbl:cuda-apis}. More
specifically, we write wrappers for these functions of interest which
request/release bandwidth lock on behalf of the GPU application before calling
the actual CUDA library functions. We compile these functions as a shared
library and use Linux' \texttt{LD\_PRELOAD}
mechanism~\cite{GregKroahHartman2004} to force the GPU application to use those
wrapper functions whenever the CUDA functions are called. In this way, we
automatically throttle CPU cores' bandwidth usage whenever real-time GPU kernels
are being executed so that the GPU kernels' memory bandwidth can be guaranteed.

A complication to the automatic GPU kernel instrumentation arises when the
application uses CUDA streams to launch multiple GPU kernels in succession in
multiple streams and then waits for those kernels to complete. In this case, the
bandwidth lock acquired by a GPU kernel launched in one stream can potentially
be released when synchronizing with a kernel launched in another stream. In our
framework, this situation is averted by keeping track of active streams and
associating bandwidth lock with individual streams instead of the entire
application whenever stream based CUDA APIs are invoked. A stream is considered
active if:
\begin{itemize}
\item A kernel or memory copy operation is launched in that stream
\item The stream has not been explicitly (using \texttt{cudaStreamSynchronize})
or implicitly (using \texttt{cudaDeviceSynchronize} or
\texttt{cudaThreadSynchronize}) synchronized with
\end{itemize}

Our framework ensures that a GPU application continues holding the bandwidth
lock as long as it has one or more active streams.

The obvious drawback of throttling CPU cores is that the CPU throughput may be
affected especially if some of the tasks on the CPU cores are memory bandwidth
intensive. In the following sub-section, we discuss the impact of throttling on
CPU throughput and present a new CPU scheduling algorithm that minimizes
throughput reduction.

\subsection{Throttle Fair CPU Scheduler}\label{sec:tfs-intro}
As described earlier in this section, BWLOCK++ uses a throttling based approach
to enforce memory bandwidth limit of CPU cores at a regular interval. Although
effective in protecting critical GPU applications in the presence of memory
intensive CPU applications, this approach runs into the risk of severely
under-utilizing the system's CPU capacity; especially in cases when there are
multiple best-effort CPU applications with different memory characteristics
running on the best-effort CPU cores. In the throttling based design, once a
core exceeds its memory bandwidth quota and gets throttled, that core cannot be
used for the remainder of the period. Let us denote the regulation period as $T$
(i.e., $T = 1ms$) and the time instant at which an offending core exceeds its
bandwidth budget as $t$. Then the wasted time due to throttling can be described
as $\delta = T - t$ and the smaller the value of $t$ (i.e., throttled earlier in
the period) the larger the penalty to the overall system throughput. The value
of $t$ depends on the rate at which a core consumes its allocated memory budget
and that in turn depends on the memory characteristics of the application
executing on that core. To maximize the overall system throughput, the value of
$\delta$ should be minimized---that is if throttling never occurs, $t \geq T
\Rightarrow \delta = 0$, or occurs late in the period, throughput reduction will
be less.

\subsubsection{Negative Feedback Effect of Throttling on CFS}
One way to reduce CPU throttling is to schedule less memory bandwidth demanding
tasks on the best-effort CPU cores while the GPU is holding the bandwidth lock.
Assuming that each best-effort CPU core has a mix of memory bandwidth intensive
and CPU intensive tasks, then scheduling the CPU intensive tasks while the GPU
is holding the lock would reduce CPU throttling or at least delay the instant at
which throttling occurs, which in turn would improve CPU throughput.
Unfortunately, Linux's default scheduler CFS~\cite{cfslinux} actually aggravates
the possibility of early and frequent throttling when used with BWLOCK++'s
throttling mechanism.

The CFS algorithm tries to allocate fair amount of CPU time among tasks by using
each task's weighted virtual runtime (i.e., weighted execution time) as the
scheduling metric. Concretely, a task $\tau_i$'s virtual runtime $V_i$
is defined as 
\begin{align}
& V_i = \frac{E_i}{W_i}
\end{align}

where $E_i$ is the actual runtime and $W_i$ is the weight of the
task. The CFS scheduler simply picks the task with the smallest
virtual runtime.

The problem with memory bandwidth throttling under CFS arises because the
virtual run-time of a memory intensive task, which gets frequently throttled,
increases more slowly than the virtual run-time of a compute intensive task
which does not get throttled. Due to this, the virtual runtime based arbitration
of CFS tends to schedule the memory intensive tasks more than the CPU intensive
tasks while bandwidth regulation is in place.

\begin{figure*}[t]
\centering
\subfigure[Example schedule under CFS with 1-msec scheduling tick]{
	\includegraphics[width=0.85\linewidth, height = 4.5cm]{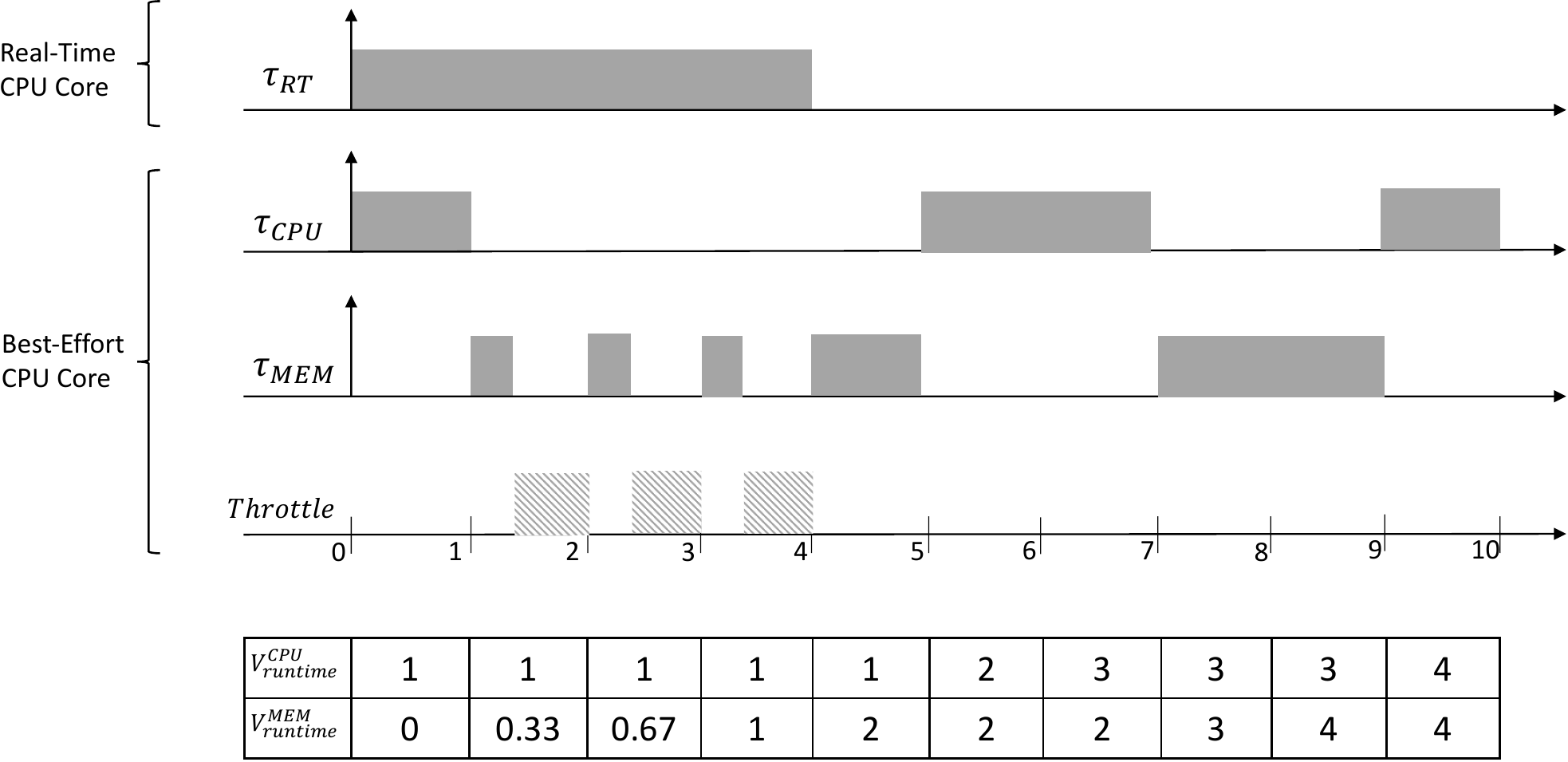}
	\label{fig:ex-cfs}
}
\hfill
\subfigure[Example schedule with zero throttling]{
	\includegraphics[width=0.85\linewidth, height = 4.5cm]{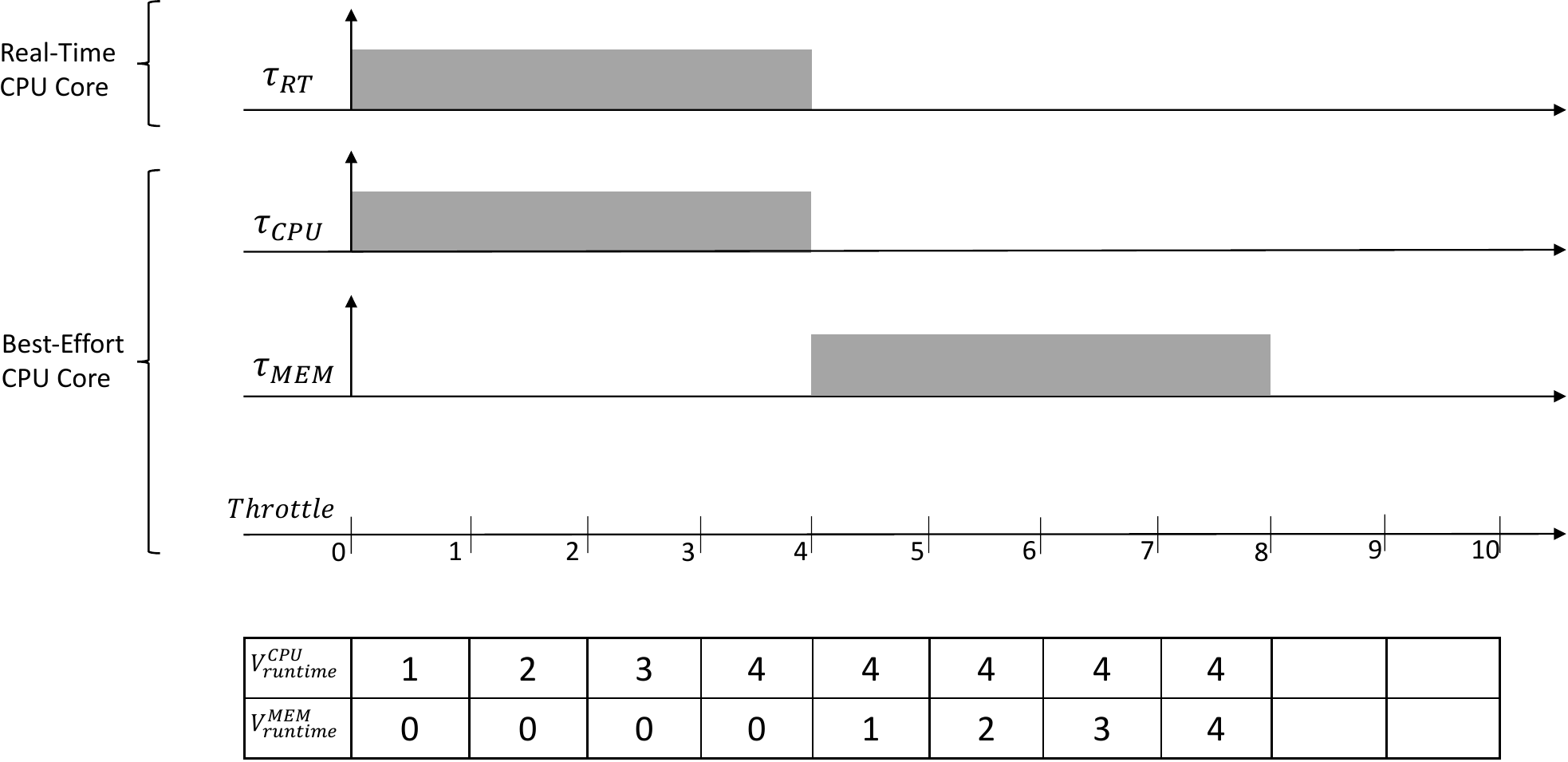}
	\label{fig:ex-ideal}
}
\hfill
\subfigure[Example schedule under TFS with $\rho = 3$]{
  	\includegraphics[width=0.85\linewidth, height = 4.5cm]{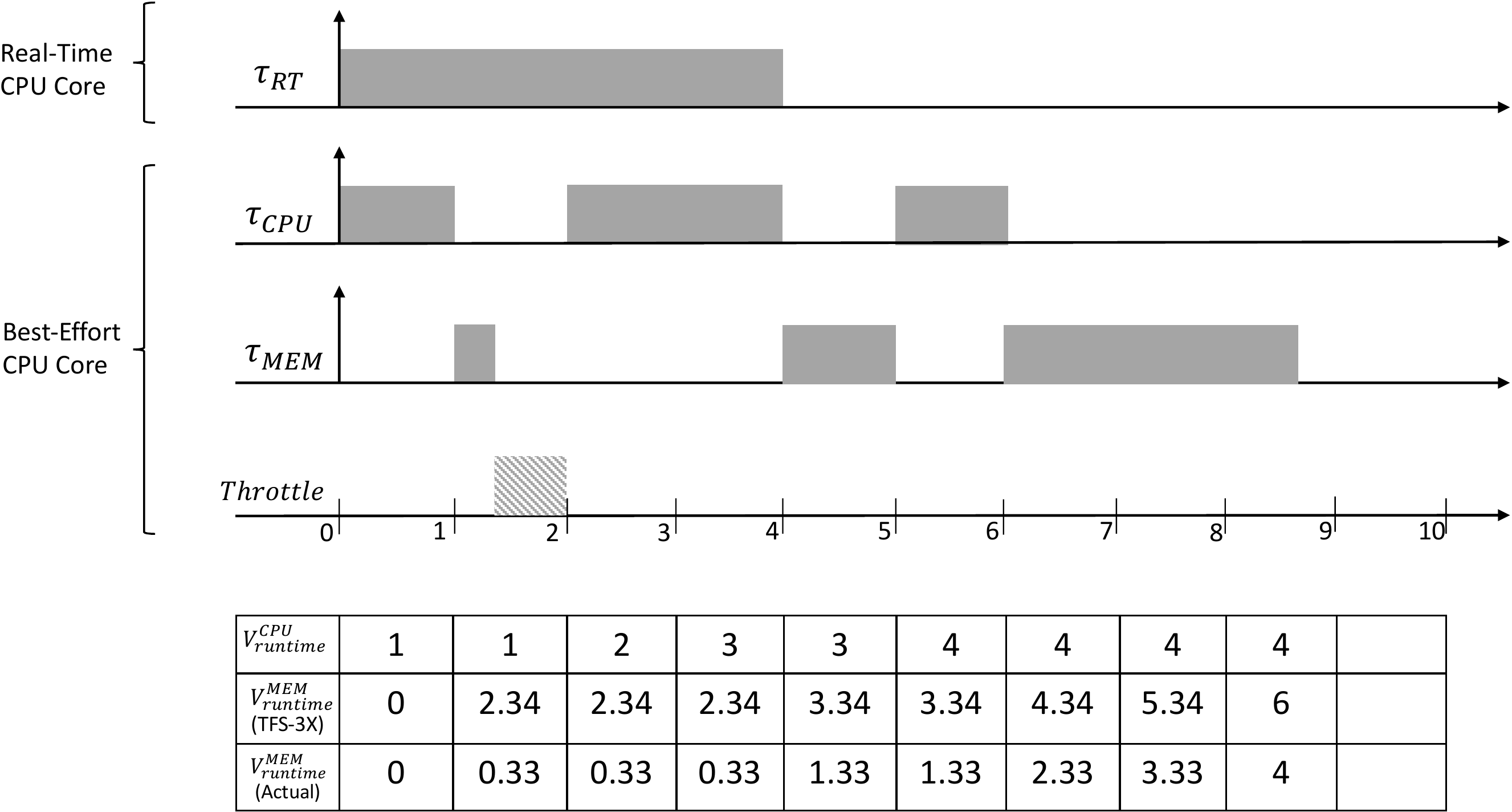}
	\label{fig:ex-tfs}
}
\caption{Example schedules under different scheduling schemes}
\label{fig:example}
\end{figure*}

\subsubsection{TFS Approach}
In order to reduce the throttling overhead while keeping the undesirable
scheduling of memory intensive tasks quantifiable, TFS modifies the throttled
task's virtual runtime to take the task's throttled duration into account.
Specifically, at each regulation period, if there exists a throttled task, we
scale the throttled duration of the task by a factor, which we call TFS
punishment factor, and add it to its virtual runtime.

Under TFS, a throttled task $\tau_i$'s virtual runtime $V_i^{new}$ at the end of
$j^{th}$ regulation period is expressed as:
\begin{align}
& V_{i}^{new} = V_{i}^{old} + \delta_{i}^{j} \times \rho \label{eq:tfs}
\end{align}

\noindent where $\delta_{i}^{j}$ is the throttled duration of $\tau_i$ in the
$j^{th}$ sampling period, and $\rho$ is the TFS punishment factor.

The more memory intensive a task is, the more likely the task get throttled in
each regulation period for a longer duration of time (i.e., higher $\delta_i$).
By adding the throttled time back to the task's virtual runtime, we make sure
that the memory intensive tasks are not favored by the scheduler. Furthermore,
by adjusting the TFS punishment factor $\rho$, we can further penalize memory
intensive tasks in favor of CPU intensive ones. This in turn reduces the amount
of throttled time and improves overall CPU utilization. On the other hand, the
memory intensive tasks will still be scheduled (albeit less frequently so)
according to the adjusted virtual runtime. Thus, no tasks will suffer
starvation.

Scheduling of tasks under TFS is fair with respect to the adjusted virtual
runtime metric but it can be considered unfair with respect to the CFS's
original virtual runtime metric. A task $\tau_i$'s ``lost'' virtual runtime
$\Delta_{i}^{TFS}$ (due to TFS's inflation) over $J$ regulation periods can be
quantified as follows:
\begin{align}
& \Delta_{i}^{TFS} = \sum_{j = 0}^J{\delta_i^{j} \times \rho}.
\end{align}

\subsubsection{Illustrative Example}
We elaborate the problem of CFS and the benefit of our TFS extension with a
concrete illustrative example.

Let us consider a small integrated CPU-GPU system, which consists of two CPU
cores and a GPU. We further assume, following our system model, that Core-1 is a
real-time core, which may use the GPU, and Core-2 is a best-effort core, which
doesn't use the GPU.
\begin{table}[h]
	\centering
	\begin{tabularx}{\linewidth}{l|p{1.3cm}|l|p{4.0cm}}
		\toprule
		Task   	& Compute Time (C) & Period (P) & Description \\
		\midrule
		$\tau_{RT}$   & 4 	& 15   & Real-time task \\
		$\tau_{MEM}$   & 4 	& N/A  & Memory intensive best-effort task \\
		$\tau_{CPU}$  & 4 	& N/A  & CPU intensive best-effort task \\
		\bottomrule
	\end{tabularx}
	\caption{Taskset for Example}
	\label{tbl:example}
\end{table}

\begin{figure*}[t]
\centering
\subfigure[CFS]{
	\includegraphics[width=0.31\textwidth, height = 5cm]{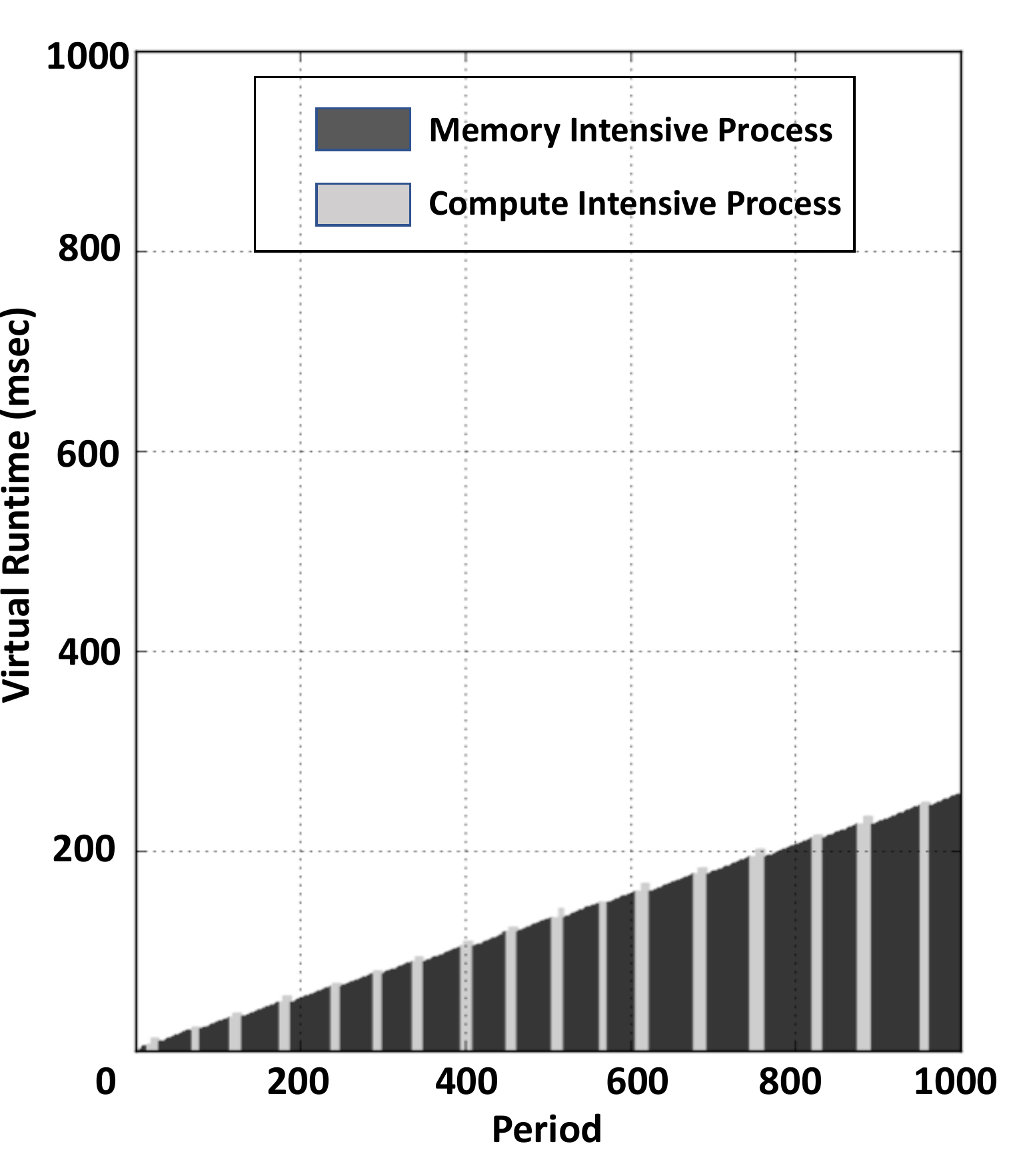}
	\label{fig:vr-thrt}
}
\hfill
\subfigure[TFS]{
	\includegraphics[width=0.31\textwidth, height = 5cm]{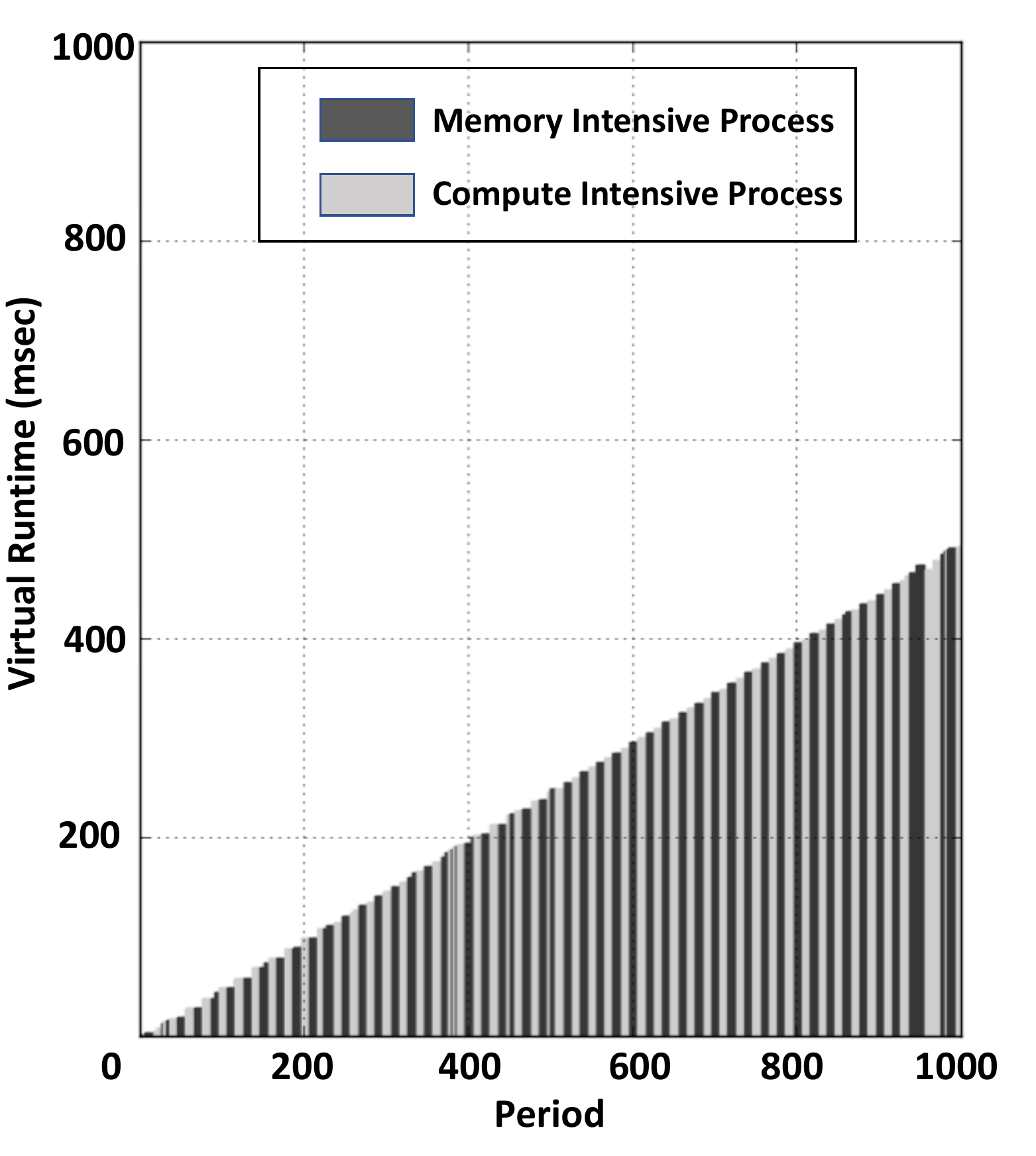}
	\label{fig:vr-tfs}
}
\hfill
\subfigure[TFS-3X]{
	\includegraphics[width=0.31\textwidth, height = 5cm]{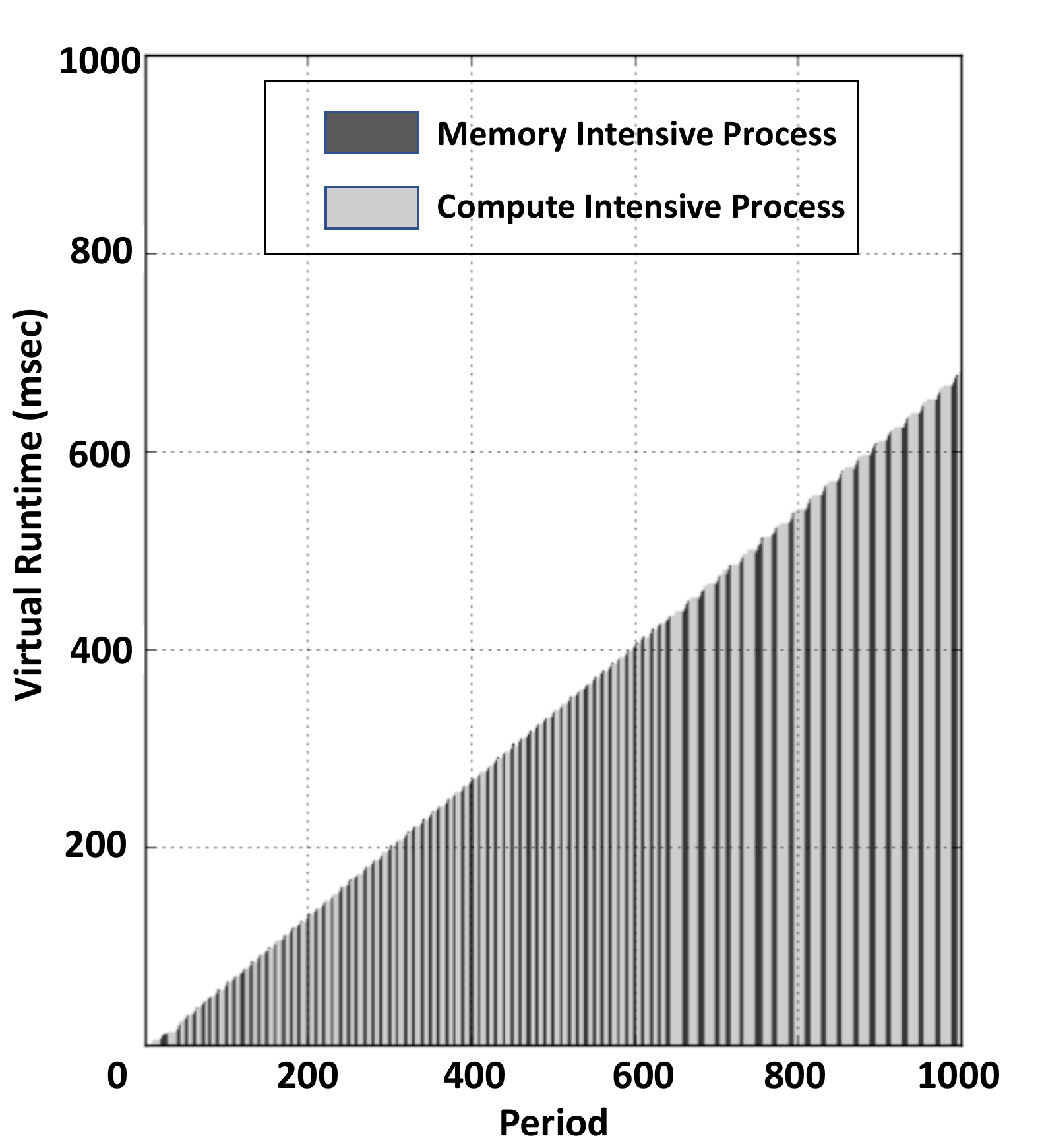}
	\label{fig:vr-tfs3}
}
\caption{Virtual runtime progress of the two synthetic tasks. One is
  cpu-intensive and the other is memory-intensive.}
\label{fig:vr-throttling-vruntime}
\end{figure*}

\begin{figure*}[h]
\centering
\subfigure[CFS]{
	\centering
	\includegraphics[width=0.31\textwidth, height = 5cm]{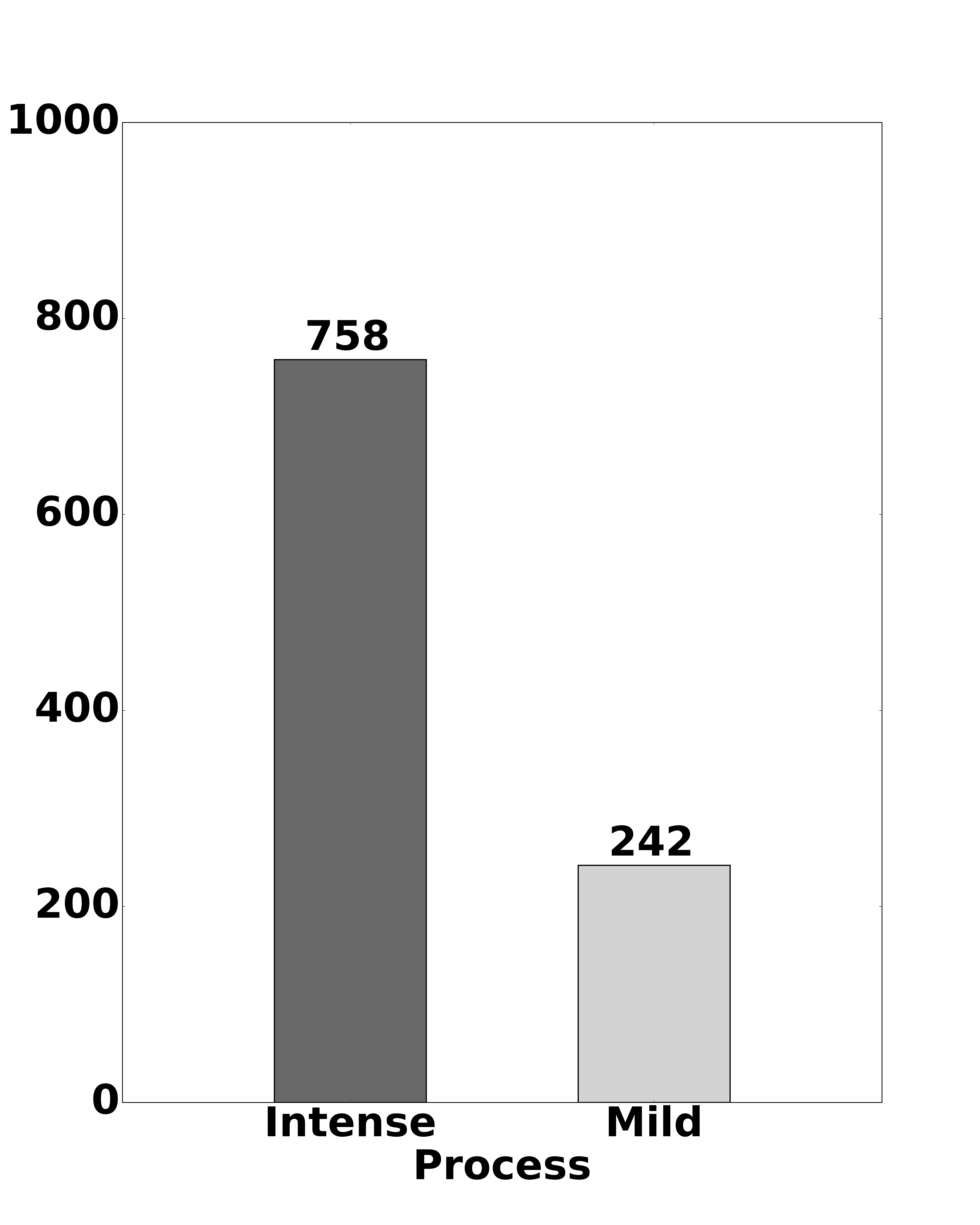}
	\label{fig:vr-thrt}
}
\hfill
\subfigure[TFS]{
	\centering
	\includegraphics[width=0.31\textwidth, height = 5cm]{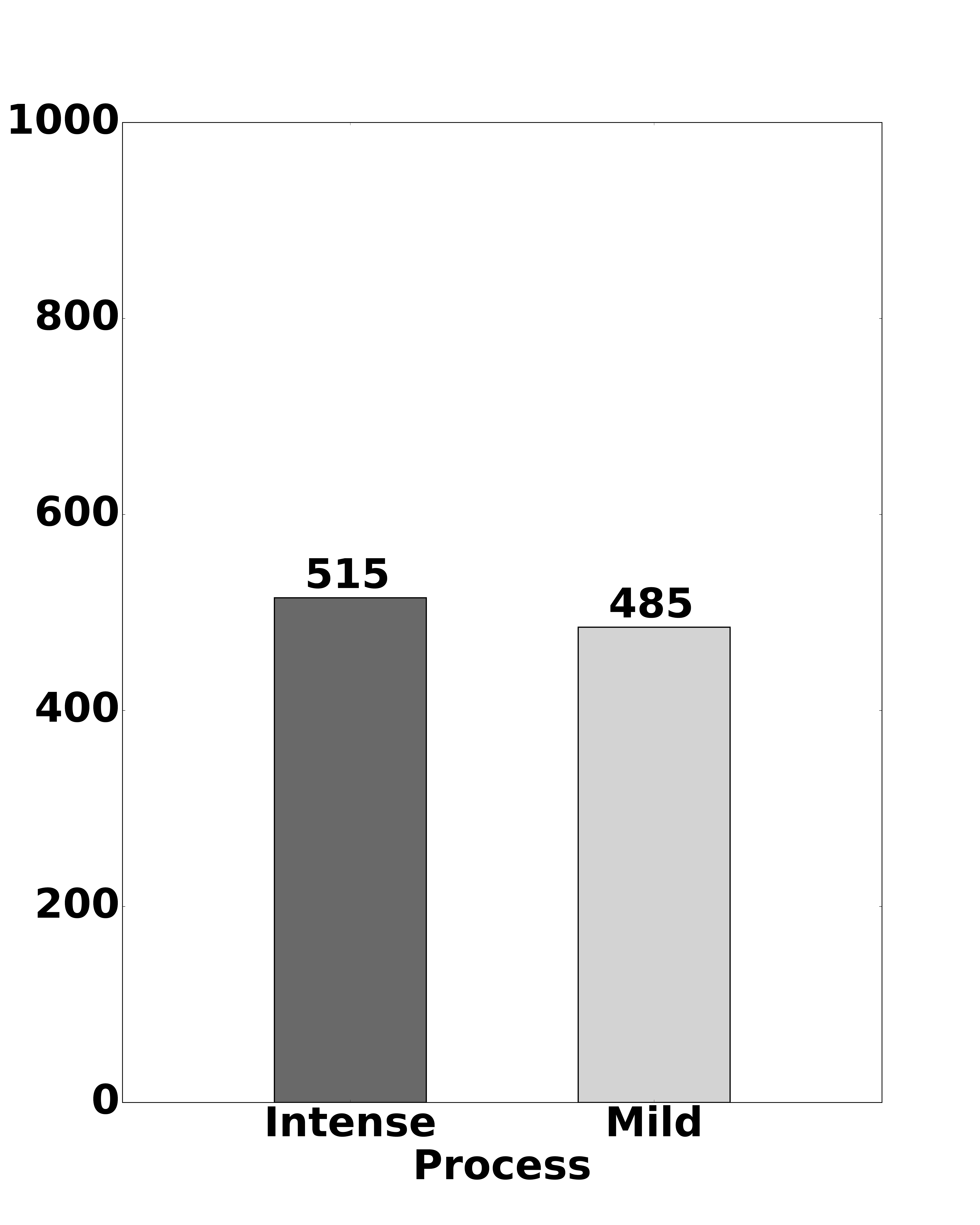}
	\label{fig:vr-tfs}
}
\hfill
\subfigure[TFS-3X]{
	\centering
	\includegraphics[width=0.31\textwidth, height = 5cm]{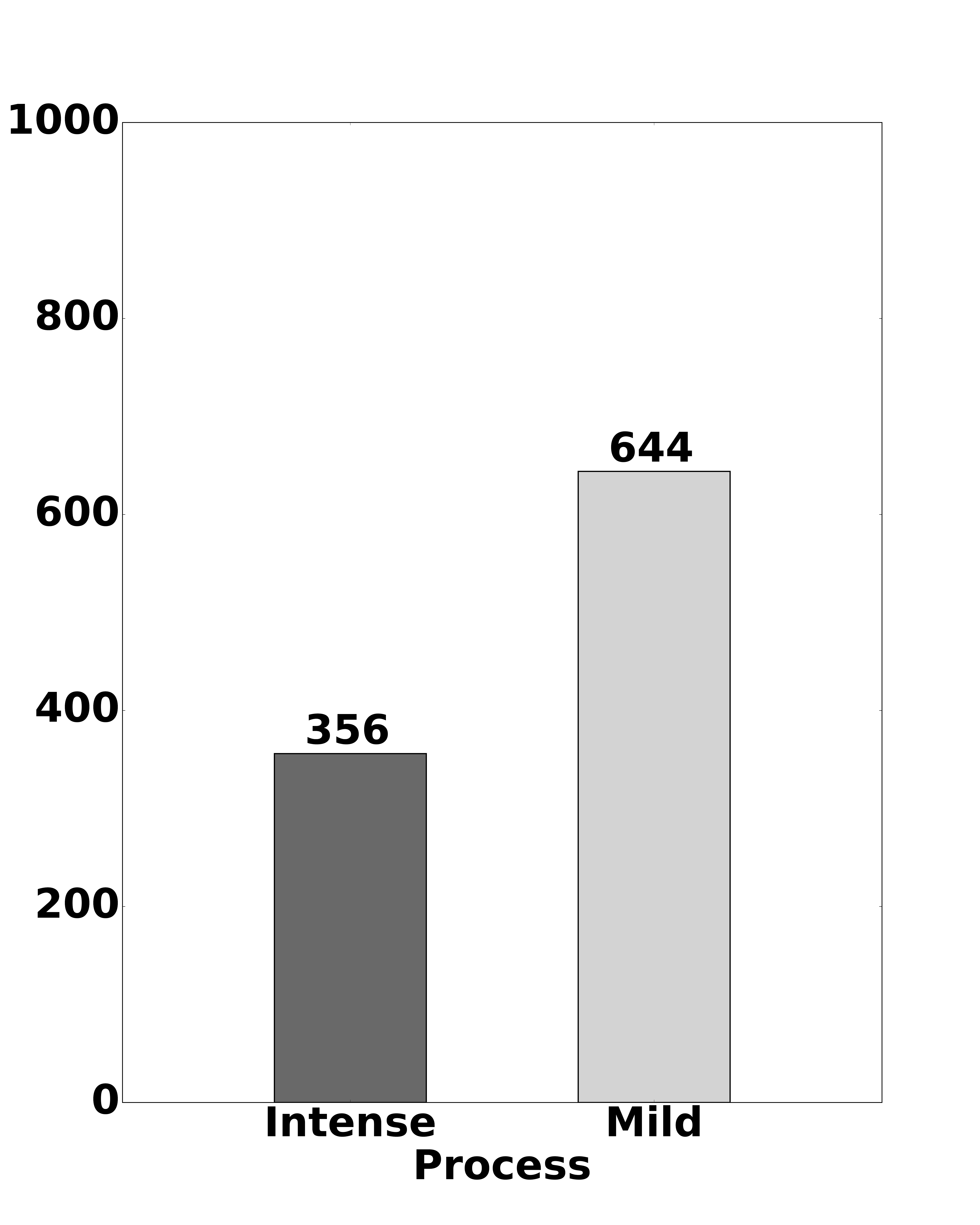}
	\label{fig:vr-tfs3}
}
\caption{The number of periods during which the two tasks are
  scheduled. 'Intense' refers to the memory-intensive task. 'Mild'
  refers to the CPU-intensive task.}
\label{fig:vr-throttling-periods}
\end{figure*}

Table~\ref{tbl:example} shows a taskset to be scheduled on the system. The
taskset is composed of a GPU using real-time task, which needs to be protected
by our framework for the entire duration of its execution; and two best-effort
tasks (of equal CFS priority), one of which is CPU intensive and the other is
memory intensive.

Figure~\ref{fig:ex-cfs} shows how the scheduling would work when CFS is used to
schedule best-effort tasks $\tau_{CPU}$ and $\tau_{MEM}$ on the best-effort core
with its memory bandwidth is throttled by our kernel-level bandwidth regulator.
Note that in this example, both OS scheduler tick timer interval and the
bandwidth regulator interval are assumed to be 1ms. At time 0, $\tau_{CPU}$ is
first scheduled. Because $\tau_{CPU}$ is CPU bound, it doesn't suffer
throttling. At time 1, the CFS schedules $\tau_{MEM}$ as its virtual runtime 0
is smaller than $\tau_{CPU}$'s virtual runtime 1. Shortly after the $\tau_{MEM}$
is scheduled, however, it gets throttled at time 1.33 as it has used the
best-effort core's allowed memory bandwidth budget for the regulation interval.
When the budget is replenished at time 2, at the beginning of the new regulation
interval, the $\tau_{MEM}$'s virtual runtime is 0.33 while $\tau_{CPU}$ is 1.
So, the CFS picks the $\tau_{MEM}$ (smaller of the two) again, which gets
throttled again. This pattern continues until the $\tau_{MEM}$'s virtual runtime
finally catches up with $\tau_{CPU}$ at time 4 by which point the best-effort
core has been throttled 66\% of time between time 1 and 4. As can be seen in
this example, CFS favors memory intensive tasks as their virtual runtimes
increase more slowly than CPU intensive ones when memory bandwidth throttling is
used.

Figure~\ref{fig:ex-ideal} shows a hypothetical schedule in which the execution
of $\tau_{MEM}$ is delayed in favor of the $\tau_{CPU}$ while $\tau_{RT}$ is
running (thus, memory bandwidth regulation is in place.) In this case, because
$\tau_{CPU}$ never exhausts the memory bandwidth budget, it never gets
throttled. As a result, the best-effort core never experiences throttling and
thus is able to achieve high throughput. While this is ideal behavior from the
perspective of throughput, it may not be ideal for the $\tau_{MEM}$ as it can
suffer starvation.

Figure~\ref{fig:ex-tfs} shows the schedule under the TFS (with a TFS punishment
factor $\rho = 3$). The TFS works identical to CFS until at time 2, when the
BWLOCK++'s periodic timer is called. At this point, the $\tau_{MEM}$'s virtual
runtime ($V^{MEM}$) is 0.33ms. However, because it has been throttled for 0.67ms
during the regulation period ($\delta = 0.67$), according to
Equation~\ref{eq:tfs}, TFS increases the task's virtual runtime to 2.34
($V^{MEM} + \delta \times \rho = 0.33 + 0.67 \times 3 = 2.34$). Because of the
increased virtual runtime, the TFS scheduler then picks $\tau_{CPU}$ as its
virtual runtime is now smaller than that of $\tau_{MEM}$ ($1 < 2.34$). Later,
when the $\tau_{CPU}$'s virtual runtime becomes 3 at time 4, the TFS scheduler
can finally re-schedule the $\tau_{MEM}$. In this manner, TFS favors CPU
intensive tasks over memory-intensive ones, while preventing starvation of the
latter. Note that TFS works at each regulation period (i.e., 1ms)
independently and thus automatically adapts to the task's changing
behavior. For example, if a task is memory intensive only for a brief
period of time, the task will be throttled only for the memory 
intensive duration, and the throttled time will be added back to the
task's virtual runtime at each 1ms regulation period. Furthermore,
even for a period when a task is throttled, the task always makes
small progress as allowed by the memory bandwidth budget for the
period. Therefore, no task suffers complete starvation for an extended
period of time.

\subsubsection{Effects of TFS using Synthetic Tasks }
We experimentally validate the effect of TFS in scheduling best-effort tasks on
a real system. In this experiment, we use two synthetic tasks: one is CPU
intensive and the other is memory-intensive. We use \emph{Bandwidth} benchmark
for both of these tasks. In order to make \emph{Bandwidth} memory intensive, we
configure its working-set size to be twice the size of LLC on our platform.
Similarly, to make \emph{Bandwidth} compute (CPU) intensive, we make its working
set size one half the size of L1 data cache in our platform. We assign these two
best-effort tasks on the same best-effort core, which is bandwidth regulated
with a 100 MB/s memory bandwidth budget.

Figure~\ref{fig:vr-throttling-vruntime} shows the virtual runtime progression
over 1000 sampling periods of the two tasks under three scheduler
configurations: CFS, TFS ($\rho = 1$), and TFS-3X ($\rho = 3$). In CFS, the
memory intensive process gets preferred by the CFS scheduler at each scheduling
instance, because its virtual run-time progresses more slowly. In TFS and
TFS-3X, however, as memory-intensive task's virtual runtime is increased,
CPU-intensive task is scheduled more frequently.

This can be seen more clearly in Figure~\ref{fig:vr-throttling-periods}, which
shows the number of periods utilized by each task on the CPU core, over the
course of one thousand sampling periods. Under CFS, out of all the sampling
periods, 75\% are utilized by the memory intensive process and only 25\% are
utilized by the compute intensive process. With TFS, the two tasks get to run in
roughly the same number of sampling periods whereas in TFS-3x, the CPU intensive
task gets to run more than the memory intensive task.

\section{Implementation}\label{sec:implementation}
In this section, we describe the implementation details of BWLOCK++.

\subsection{BWLOCK++ System Call}
\begin{algorithm}[t]
\SetAlgoLined
\DontPrintSemicolon
\SetKwInOut{Input}{Input}
\SetKwInOut{Result}{Result}
\SetKwFunction{func}{sys\_bwlock}
\SetKwProg{algo}{syscall}{}{}
\Input{Bandwidth lock value \texttt{(bw\_val)}}
\Result{Current process on RT core acquires/releases bandwidth lock and has its
priority boosted/restored}
\nonl\;
\algo{\func{bw\_val}}{
\If{smp\_processor\_id () == \texttt{RT\_CORE\_ID} $\land$ rt\_task (current)}{
	rt\_core\_data $:=$ get\_rt\_core\_data ()\;
	rt\_core\_data $\rightarrow$ current\_task $:=$ current\;
	\uIf{bw\_val $\geq$ 1}{
		current $\rightarrow$ bwlock\_val $:=$ 1\;
		current $\rightarrow$ bw\_old\_priority $:=$ current $\rightarrow$ rt\_priority\;
		current $\rightarrow$ rt\_priority $:=$ \texttt{MAX\_USER\_RT\_PRIO} - 1\;
	} \Else {
		current $\rightarrow$ bwlock\_val $:=$ 0\;
		current $\rightarrow$ rt\_priority $:=$ current $\rightarrow$ bw\_old\_priority\;
	}
}
}
\KwRet;
\caption{BWLOCK++ System Call}
\label{alg:syscall}
\end{algorithm}

We add a new system call \texttt{sys\_bwlock} in Linux kernel 4.4.38. The
system call serves two purposes. 1) It acquires or releases the memory bandwidth
lock on behalf of the currently running task on the real-time core; and 2) it
implements a priority-ceiling protocol, which boosts the calling task's priority
to the system's ceiling priority, to prevent preemption. We introduce two new
integer fields, \texttt{bwlock\_val}, \texttt{bw\_old\_priority}, in the task
control block: \texttt{bwlock\_val} stores the current status of the memory
bandwidth lock and \texttt{bw\_old\_priority} keeps track of the original
real-time priority of the task while it is holding the bandwidth lock.

Algorithm~\ref{alg:syscall} shows the implementation of the system call. To
acquire the memory bandwidth lock, the system call must be invoked from the
real-time system core and the task currently scheduled on the real-time core
must have a real-time priority (\emph{line 2}). At the time of acquisition of
bandwidth lock, the priority of the calling task, which is tracked by the
globally accessible \texttt{current} pointer in Linux kernel, is raised to the
maximum allowed real-time priority value (the ceiling priority) for any
user-space task to prevent preemption (\emph{line 7}). The real-time priority
value of the the task is restored to its original priority value when the
bandwidth lock is released (\emph{line 10}). In this manner, the system call
updates the state of the currently scheduled real-time task on the real-time
system core, which is then used by the memory bandwidth regulator on best-effort
cores to enforce memory usage thresholds, as explained in the following
subsection.

\subsection{Per-Core Memory Bandwidth Regulator}
\begin{algorithm}[t]
\SetAlgoLined
\DontPrintSemicolon
\SetKwInOut{Input}{Input}
\SetKwInOut{Result}{Result}
\SetKwFunction{funcA}{periodic\_interrupt\_handler}
\SetKwFunction{funcB}{pmc\_overflow\_handler}
\Input{Data structure containing core private information \texttt{(core\_data)}}
\Result{Memory usage threshold gets set and enforced for the core at hand for
the current regulation period. Also TFS scaling gets applied to the currently scheduled task}
\nonl\;
\SetKwProg{proc}{procedure}{}{}
\proc{\funcA{core\_data}}{
\If{core\_is\_throttled (core\_data $\rightarrow$ core\_id) == TRUE}{
	unthrottle\_core (core\_data $\rightarrow$ core\_id)\;
	record\_throttling\_end\_time (core\_data $\rightarrow$ current\_task)\;
	scale\_virtual\_runtime (core\_data $\rightarrow$ current\_task)\;
}
rt\_core\_data $:=$ get\_rt\_core\_data ()\;
\uIf{rt\_core\_data $\rightarrow$ current\_task $\rightarrow$ bwlock\_val == 1}{
	core\_data $\rightarrow$ new\_budget := rt\_core\_data $\rightarrow$ throttle\_budget\;
} \Else {
	core\_data $\rightarrow$ new\_budget := \texttt{MAX\_BANDWIDTH\_BUDGET}\;
}
program\_pmc (core\_data $\rightarrow$ new\_budget)\;
}
\KwRet;

\nonl\;
\proc{\funcB{core\_data}}{
	record\_throttling\_start\_time (core\_data $\rightarrow$ current\_task)\;
	throttle\_core (core\_data $\rightarrow$ core\_id)\;
}
\KwRet;
\caption{Memory Bandwidth Regulator}
\label{alg:regulator}
\end{algorithm}

The per-core memory bandwidth regulator is composed of a periodic timer
interrupt handler and a performance monitoring counter (PMC) overflow interrupt
handler. Algorithm~\ref{alg:regulator} shows the implementation of the memory
bandwidth regulator.

The periodic timer interrupt handler is invoked at a periodic interval
(currently every 1 msec) using a high resolution timer in each best-effort core.
The timer handler begins a new bandwidth lock regulation period and performs the
following operations:
\begin{itemize}
\item Unthrottle the core if it was throttled in the last regulation period (\emph{line 3})
\item Scale the virtual runtime of the task currently scheduled on the core
based on the throttling time in the last period and the TFS punishment factor
(\emph{line 4-5})
\item Determine the new memory usage budget based on the bandwidth lock status
of the task currently scheduled on the real-time system core
(\emph{line 7-12})
\item Program the performance monitoring counter on the core based on the new
memory usage budget for the current regulation period (\emph{line 13}). We use
the \emph{L2D\_CACHE\_REFILL} event for measuring the memory bandwidth traffic
in ARM Cortex-A57 processor core
\end{itemize}

The PMC overflow interrupt occurs when the core at hand exceeds its memory usage
budget in the current regulation period. The interrupt handler prevents further
memory transactions from this core by scheduling a high priority idle kernel
thread on it for the remainder of the regulation period (\emph{line 17}).

\section{Evaluation}\label{sec:eval}
In this section, we present the experimental evaluation results of
BWLOCK++. 

\subsection{Setup}
We evaluate BWLOCK++ on NVIDIA Jetson TX2 platform. We use the Linux kernel
version 4.4.38, which is patched with the changes required to support BWLOCK++.
The CUDA runtime library version installed on the platform is 8.0, which is the
latest version available for Jetson TX2 at the time of writing. In all our
experiments, we place the platform in maximum performance mode by maximizing GPU
and memory clock frequencies and disabling the dynamic frequency scaling of CPU
cores. We also shutdown the graphical user interface and disable the network
manager to avoid run to run variation in the experiments. As per our system
model, we designate the Core-0 in our system as real-time core. The remaining
cores execute best-effort tasks only. All the tasks are statically assigned to
their respective cores during the experiment.
While NVIDIA Jetson TX2 platform contains two CPU islands, a quad-core
Cortex-A57 and a dual-core Denver, we only use the Cortex-A57 island
for our evaluation and leave the Denver island off because we were
unable to find publicly available documentation regarding the Denver
cores' hardware performance counters, which is needed to implement throttling.
In order to evaluate BWLOCK++, we use six benchmarks from parboil suite which
are listed as memory bandwidth sensitive in \cite{parboil}.

\subsection{Effect of Memory Bandwidth Contention}\label{sec:eval-contention}
In this experiment, we investigate the effect of memory bandwidth contention due
to co-scheduled memory intensive CPU applications on the evaluated GPU kernels. 

\begin{table*}[t]
	\centering
		\begin{tabularx}{\textwidth}{p{2.15cm} | p{2.15cm} | p{2.15cm} | p{2.15cm} | p{2.15cm} | p{2.15cm} | p{2.15cm}}
		\toprule
		\multirow{2}{*}{Benchmark} & \multirow{2}{*}{Dataset} & Copy Amount & \multicolumn{4}{c}{Timing Breakdown (msec)} \\
                \cline{4-7}
                	&		  & (KBytes)   & Kernel ($G^e$) &  Copy ($G^m$) & Compute ($C$) & Total ($E$) \\
		\hline
		histo	& Large   	  &	5226   & 83409 	 & 18	& 0	& 83428	 \\
		sad	& Large	   	  &	709655 & 152	 & 654	& 53	& 861	 \\
		bfs	& 1M 		  &	62453  & 174	 & 72	& 0	& 246	 \\
		spmv	& Large	   	  &	30138  & 69	 & 51	& 10	& 131	 \\
		stencil & Default 	  &	196608 & 749	 & 129	& 9	& 888	 \\
		lbm	& Long   	  &	379200 & 43717   & 358	& 2004	& 46080  \\
		\bottomrule
		\end{tabularx}
	\caption{GPU execution time breakdown of selected benchmarks}
	\label{tbl:profile}
\end{table*}

\begin{figure}[t]
  \centering
  \includegraphics[width=0.9\linewidth, height=7cm] {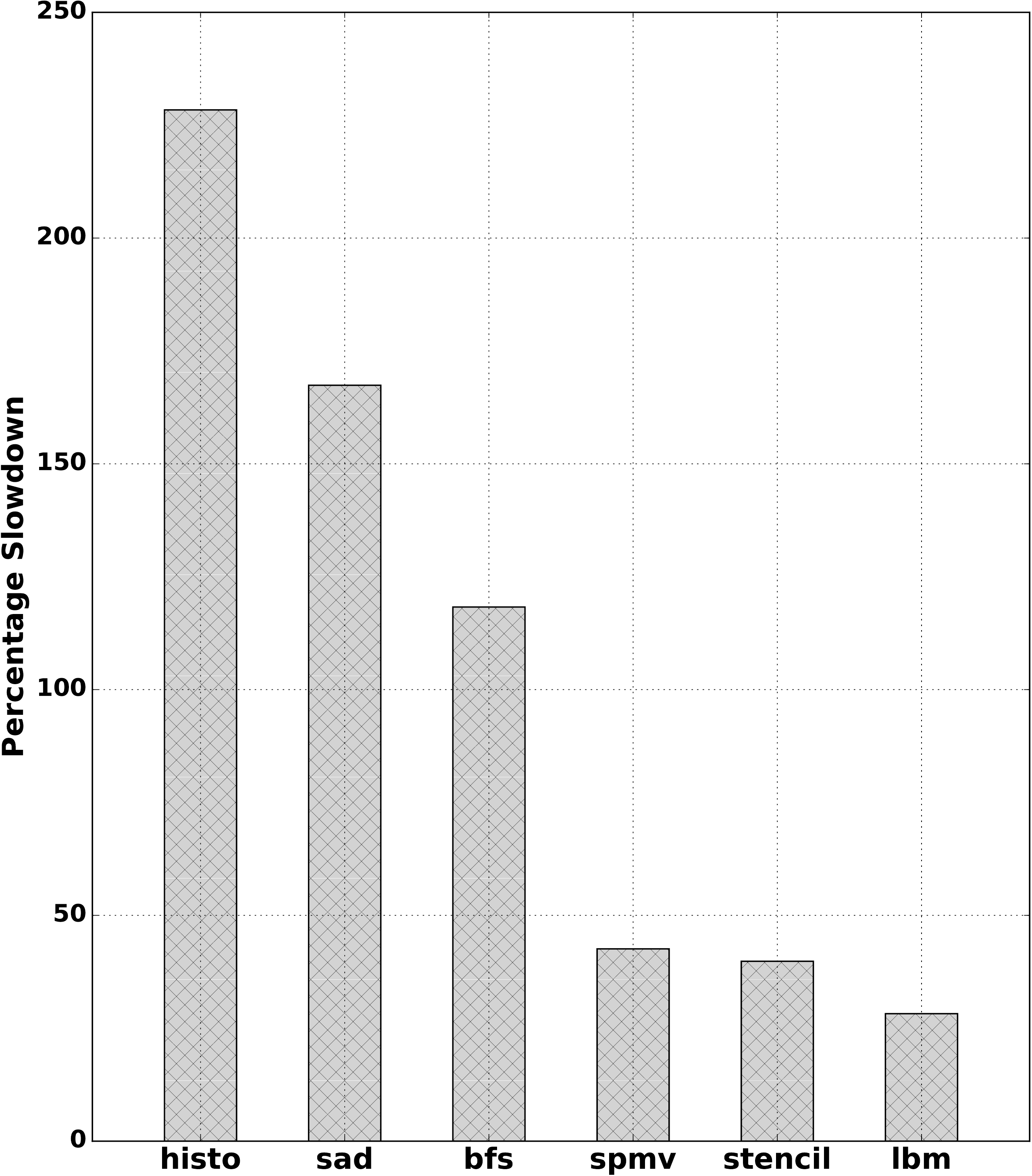}
  \caption{Slowdown of the total execution time of GPU benchmarks due to three
  \emph{Bandwidth} corunners}
  \label{fig:eval-slowdown}
\end{figure}

First, we measure the execution time of each GPU benchmark in isolation. From
this experiment, we record the GPU kernel execution time ($G^e$), memory copy
time for GPU kernels ($G^m$) and CPU compute time ($C$) for each benchmark. The
data collected is shown in Table~\ref{tbl:profile}. We then repeat the
experiment after co-scheduling three instances of a memory intensive CPU
application as co-runners. We use the \emph{Bandwidth} benchmark from the
IsolBench suite \cite{valsan2016} as the memory intensive CPU benchmark, which
updates a big 1-D array sequentially. The sequential write access pattern of the
benchmark is known to cause worst-case interference on several multicore
platforms~\cite{Valsan2017}.

The results of this experiment are shown in Figure~\ref{fig:eval-slowdown} and
they demonstrate how much the total execution time of GPU benchmarks ($E = G^e +
G^m + C$) suffers from memory bandwidth contention due to the co-scheduled CPU
applications.

From Figure~\ref{fig:eval-slowdown}, it can be seen that the worst case
slowdown, in case of \emph{histo} benchmark, is more than 250\%. Similarly, for
SAD benchmark, the worst case slowdown is more than 150\%. For all other
benchmarks, the slowdown is non-zero and can be significant in affecting the
real-time performance. These results clearly show the danger of uncontrolled
memory bandwidth sharing in an integrated CPU-GPU architecture as GPU kernels
may potentially suffer severe interference from co-scheduled CPU applications.
In the following experiment, we investigate how this problem can be addressed by
using BWLOCK++.

\subsection{Determining Memory Bandwidth Threshold}\label{sec:eval-thresh}
In order to apply BWLOCK++, we first need to determine safe memory budget that
can be given to the best-effort CPU cores in the presence of GPU applications.
However, an appropriate threshold value may vary depending on the
characteristics of individual GPU applications. If the threshold value is set
too high, then it may not be able to protect the performance of the GPU
application. On the other hand, if the threshold value is set too low, then the
CPU applications will be throttled more often and that would result in
significant CPU capacity loss.

We calculate the safe memory budget for best-effort CPU cores by observing the
trend of slowdown of the total execution time of GPU application as the allowed
memory usage threshold of CPU co-runners is varied. We start with a threshold
value of 1-GB/s for each best-effort CPU core. We then continue reducing the
threshold value for best-effort cores by half and measure the impact of this
reduction on the slowdown of execution time ($E$) of the benchmark.
\begin{figure}[h]
  \centering
  \includegraphics[width=0.9\linewidth, height=7cm] {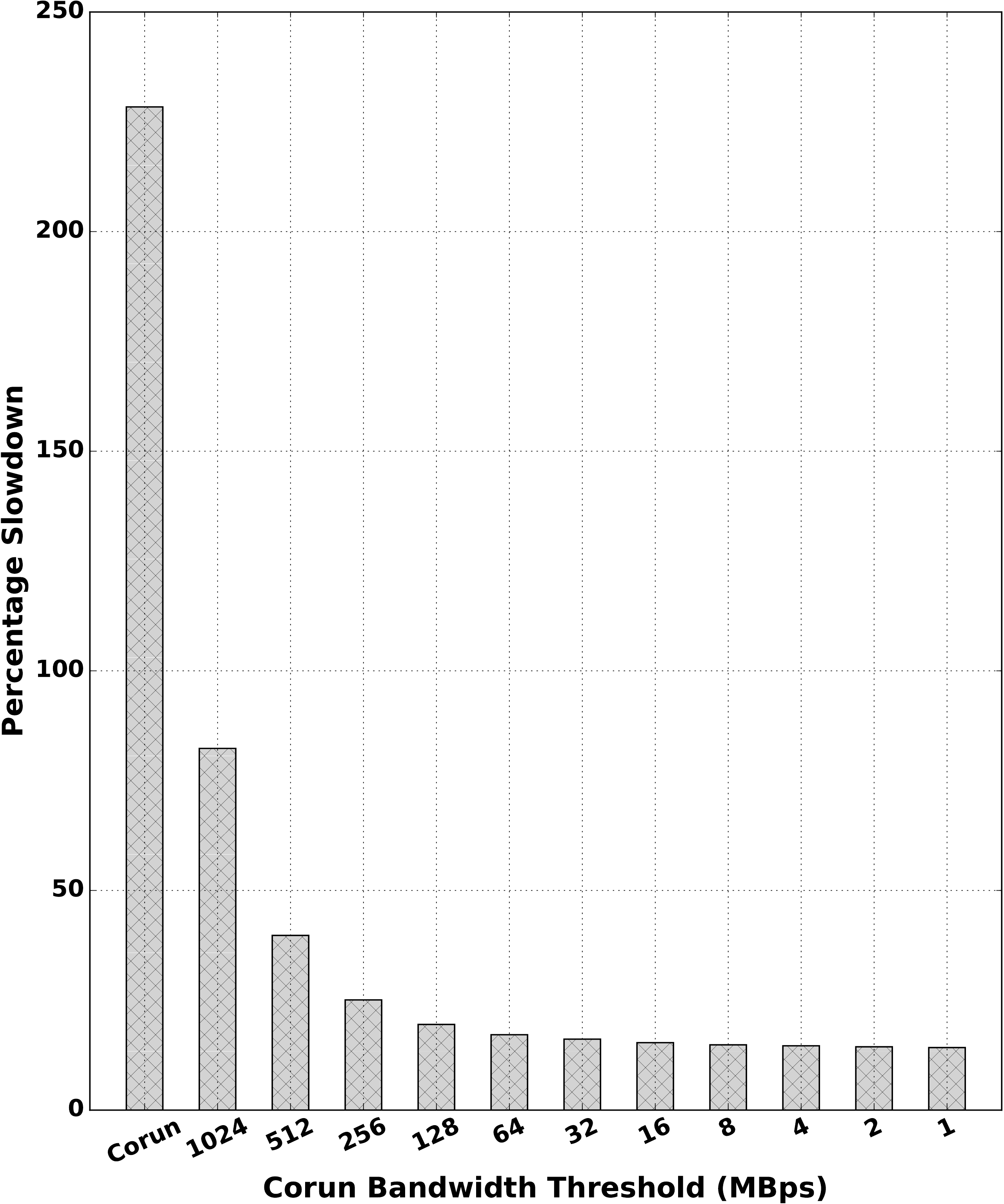}
  \caption{Effect of corun bandwidth threshold on the execution time of histo benchmark}
  \label{fig:threshold}
\end{figure}

We calculate the safe memory budget for best-effort CPU cores by observing the
trend of slowdown of the total execution time of GPU application as the allowed
memory usage threshold of CPU co-runners is varied. We start with a threshold
value of 1-GB/s for each best-effort CPU core. We then continue reducing the
threshold value for best-effort cores by half and measure the impact of this
reduction on the slowdown of execution time ($E$) of the benchmark.

\subsection{Effect of BWLOCK++}\label{sec:eval-bwlock}
In this experiment, we evaluate the performance of BWLOCK++. Specifically, we
record the corun execution of GPU benchmarks with the automatic instrumentation
of BWLOCK++. We call this scenario \emph{BW-Locked-Auto}. We compare the
performance under \emph{BW-Locked-Auto} against the \emph{Solo} and \emph{Corun}
execution of the GPU benchmarks which represent the measured execution times in
isolation and together with three co-scheduled memory intensive CPU
applications, respectively.
\begin{figure*}[t]
  \centering
  \includegraphics[width=1.0\textwidth, height=6cm] {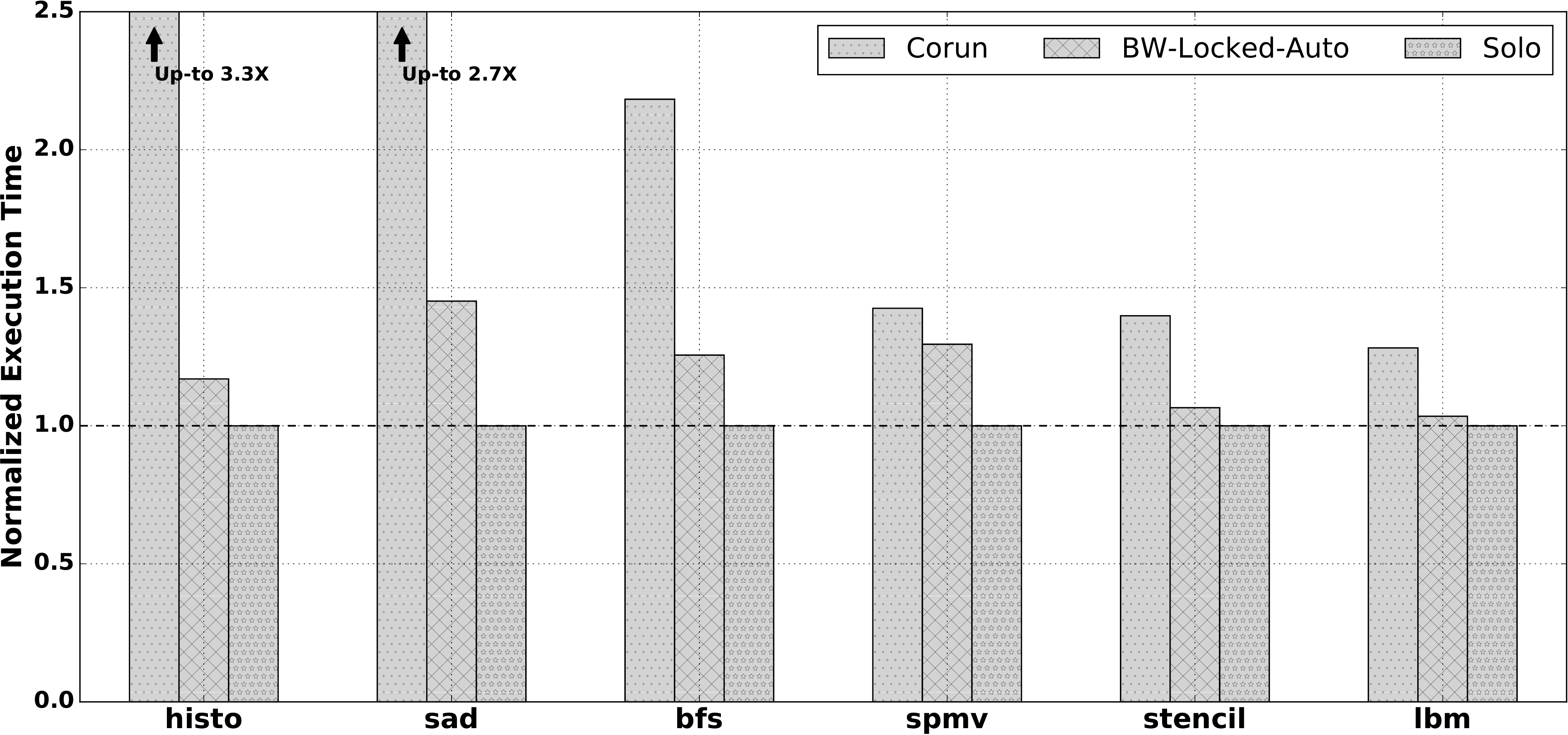}
  \caption{BWLOCK++ Evaluation Results}
  \label{fig:evaluation}
\end{figure*}

\begin{figure*}[h]
  \centering
  \includegraphics[width=1.0\textwidth, height=6cm] {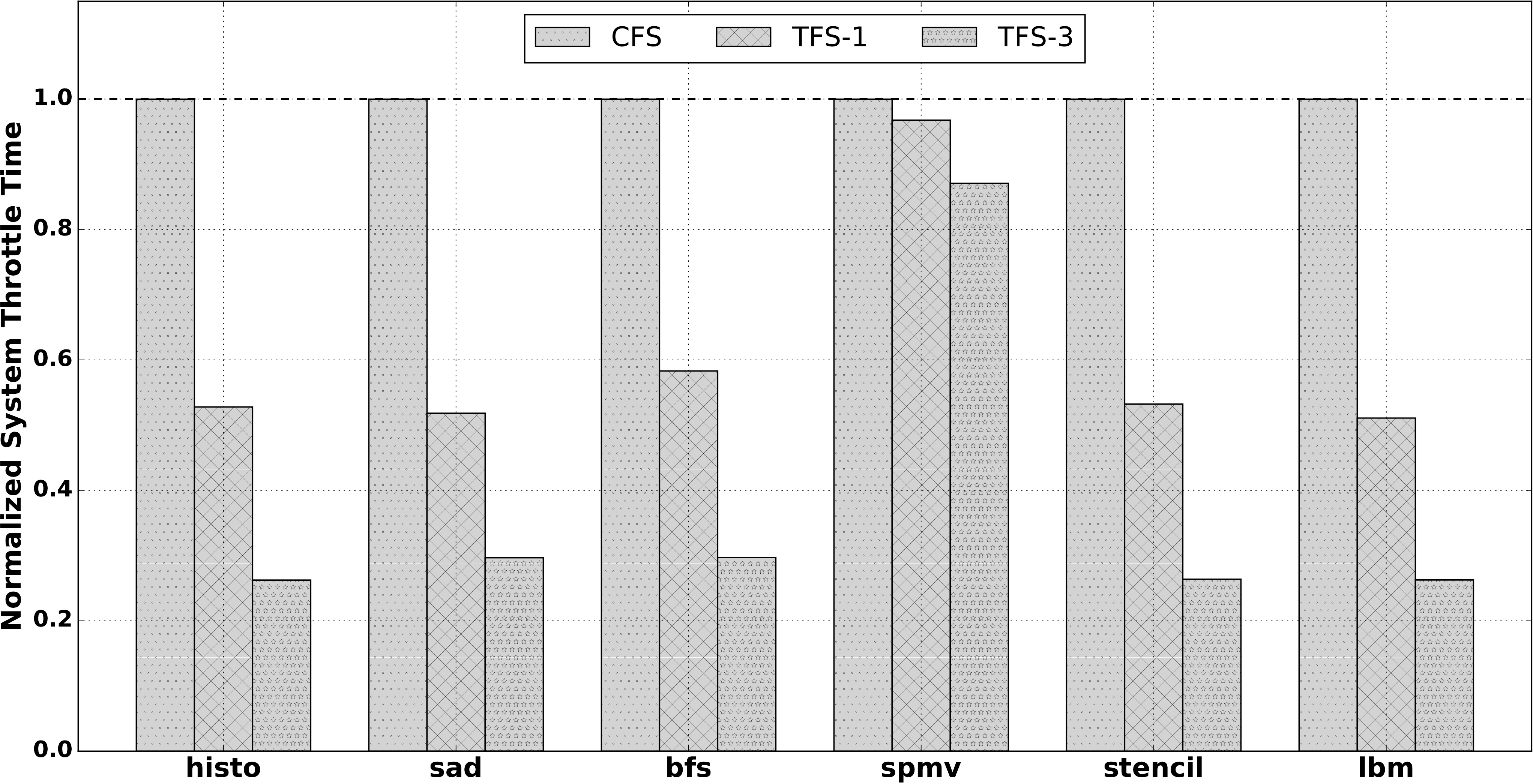}
  \caption{Comparison of total system throttle time under different scheduling schemes}
  \label{fig:tfs-eval}
\end{figure*}

To get the data-points for \emph{BW-Locked-Auto}, we configure BWLOCK++
according to the allowed memory usage threshold of the benchmark at hand and use
our dynamic GPU kernel instrumentation mechanism to launch the benchmark in the
presence of three \emph{Bandwidth} benchmark instances (write memory access
pattern) as CPU co-runners. The results of this experiment are plotted in
Figure~\ref{fig:evaluation}. In Figure~\ref{fig:evaluation}, we plot the total
execution time of each benchmark for the above mentioned scenarios. All the time
values are normalized with respect to the total execution time ($E_{solo} =
C_{solo} + G^e_{solo} + G^m_{solo}$) of the benchmark in isolation. As can be
seen from this figure, execution under \emph{BW-Locked-Auto} incurs
significantly less slowdown of the total execution time of GPU benchmarks due to
reduction of both GPU kernel execution time and memory copy operation time.

\subsection {Throughput improvement with TFS}\label{sec:tfs-eval}
As explained in Section~\ref{sec:tfs-intro}, throttling under CFS results in
significant system throughput reduction. In order to illustrate this, we conduct
an experiment in which the GPU benchmarks are executed with six CPU co-runners.
Each CPU core, apart from the one executing the GPU benchmark, has a memory
intensive application and a compute intensive application scheduled on it. For
both of these applications, we use the \emph{Bandwidth} benchmark with different
working set sizes. In order to make \emph{Bandwidth} memory intensive, we
configure its working set size to be twice the size of LLC on our evaluation
platform. Similarly for compute intensive case, we configure the working set
size of \emph{Bandwidth} to be half of the L1-data cache size. We record the
total system throttle time statistics with BWLOCK++ for all the GPU benchmarks.
The total system throttle time is the sum of throttle time across all system
cores. We then repeat the experiment with our Throttle Fair Scheduling scheme.
In \emph{TFS-1}, we configure the TFS punishment factor as one for the memory
intensive threads and in \emph{TFS-3}, we set this factor to three. We plot the
normalized total system throttle time for all the scheduling schemes and present
them in Figure~\ref{fig:tfs-eval}. It can be seen that TFS results in
significantly less system throttling (On average, $39\%$ with \emph{TFS-1} and
$62\%$ with \emph{TFS-3}) as compared to CFS.

\subsection {Overhead due to BWLOCK++}\label{sec:bw-overhead}
The overhead incurred by real-time GPU applications due to BWLOCK++ comes from
the following sources:
\begin{itemize}
\item \texttt{LD\_PRELOAD} overhead for CUDA API instrumentation
\item Overhead due to BWLOCK++ system call
\end{itemize}

The overhead due to \texttt{LD\_PRELOAD} is negligible since we cache CUDA API
symbols for all the instrumented functions inside our shared library; after
searching for them only once through the dynamic linker. We calculate the
overhead incurred due to BWLOCK++ system call by executing the system call one
million times and taking the average value. In NVIDIA Jetson TX2, the average
overhead due to each BWLOCK++ system call is $1.84 usec$. Finally, we
experimentally determine the overhead value for all the evaluated benchmarks by
running the benchmark in isolation with and without BWLOCK++. Our experiment
shows that for all the evaluated benchmarks, the total overhead due to BWLOCK++
is less than $1\%$ of the total solo execution time of the benchmark.

\section{Schedulability Analysis}\label{sec:analysis}
As we limit the scheduling of real-time tasks on a single real-time
core, our system can be analyzed using the classical unicore based response time
analysis for preemptive fixed priority scheduling with
blocking~\cite{Audsley93RTA}, because we model each GPU execution segment as a
critical section, which is protected by acquiring and releasing the bandwidth
lock. The bandwidth lock serializes GPU execution and regulates memory bandwidth
consumption of co-scheduled best-effort CPU tasks. The bandwidth lock implements
the standard priority ceiling protocol~\cite{sha1990priority}, which boosts the
priority of the lock holding task (i.e., the task executing a GPU kernel) to the
ceiling priority of the lock, which is the highest real-time priority of the
system, so as to prevent preemption. With this constraint, a real-time task
$\tau_i$'s response time is expressed as:
\begin{align}
& R_i^{n+1} = E_i + B_i + \sum_{\forall j \in
 hp(i)}{\left\lceil\frac{R_i^n}{P_{j}}\right\rceil}E_{j}
\label{eq:rta}
\end{align}

\noindent where $hp(i)$ represents the set of higher priority tasks than
$\tau_i$ and $B_i$ is the longest GPU kernel or copy
duration---protected by the memory bandwidth lock---of one of the lower priority
tasks.

The benefit of BWLOCK++ lies in the reduction of worst-case GPU kernel execution
or GPU memory copy interval of real-time tasks (which would in turn reduce $E_i$
and $B_i$ terms in Equation~\ref{eq:rta}). As shown in
Section~\ref{sec:eval-contention}, without BWLOCK++, GPU execution of a task can
suffer severe slowdown (up to $230\%$ slowdown in our evaluation), which would
result in pessimistic WCET estimation for GPU kernel and copy execution times,
hampering schedulability of the system. BWLOCK++ helps reduce pessimism
of GPU execution time estimation and thus improves schedulability.

\section{Discussion}\label{sec:discussion}
Our approach has following limitations. First, we assume that all real-time
tasks are scheduled on a single dedicated real-time core while the rest of the
cores only schedule best-effort tasks. In addition, we assume only real-time
tasks can utilize the GPU while best-effort tasks cannot. While restrictive,
recall that scheduling multiple GPU using real-time tasks on a single dedicated
real-time core does not necessarily reduce GPU utilization because multiple GPU
kernels from different tasks (processes) are serialized at the GPU hardware
anyway~\cite{nathan2017a} as we already discussed in Section~\ref{sec:model}.
Also, due to the capacity limitation of embedded GPUs, it is expected that a few
GPU using real-time task can easily achieve high GPU utilization in practice.
We claim that our approach is practically useful for situations where a small
number of GPU accelerated tasks are critical, for example, a vision-based
automatic braking system.

Second, we assume that GPU applications are given a priori and they can be
profiled in advance so that we can determine proper memory bandwidth threshold
values. If this assumption cannot be satisfied, an alternative solution is to
use a single threshold value for all GPU applications, which eliminates the need
of profiling. But the downside is that it may lower the CPU throughput because
the memory bandwidth threshold must be conservatively set to cover all types of
GPU applications.

\section{Conclusion}\label{sec:conclusion}
In this paper, we presented BWLOCK++, a software based mechanism for protecting
the performance of GPU kernels on platforms with integrated CPU-GPU
architectures.

BWLOCK++ automatically instruments GPU applications at run-time and inserts a
memory bandwidth lock, which throttles memory bandwidth usage of the CPU cores
to protect performance of GPU kernels. We identified a side effect of memory
bandwidth throttling on the performance of Linux default scheduler CFS, which
results in the reduction of overall system throughput. In order to solve the
problem, we proposed a modification to CFS, which we call Throttle Fair
Scheduling (TFS) algorithm. Our evaluation results have shown that BWLOCK++
effectively protects the performance of GPU kernels from memory intensive CPU
co-runners. Also, the results showed that TFS improves system throughput,
compared to CFS, while protecting critical GPU kernels. In the future, we plan
to evaluate BWLOCK++ on other integrated CPU-GPU architecture based platforms.
We also plan to extend BWLOCK++ not only to protect critical GPU tasks but also
to protect critical CPU tasks.

%% \section*{Appendix} \label{appendix}

\section*{Acknowledgements} \label{acknowledge}
This research is partly supported by NSF CNS 1718880.

\balance
\bibliographystyle{unsrt}
\bibliography{ms}

\end{document}